\documentclass[aps,pra,amsmath,amssymb,twocolumn,noshowpacs,superscriptaddress]{revtex4-2}  
\usepackage{graphicx}  
\usepackage{dcolumn}   
\usepackage{bm}        
\usepackage{amssymb}   
\usepackage{amsmath}
\usepackage{mathtools}
\usepackage{color}
\usepackage{url}
\usepackage[free-standing-units=true]{siunitx} 
\usepackage[colorlinks=true, pdfstartview=FitV, linkcolor=blue, citecolor=blue, urlcolor=blue]{hyperref} 
\usepackage{epstopdf}
\usepackage{amsthm}
\usepackage{amsfonts}
\usepackage{siunitx}

\usepackage{appendix} 

\newcommand{\eff}{\mathrm{eff}}

\newcommand{\kb}{k_\mathrm{B}}
\renewcommand{\min}{\mathrm{min}}
\renewcommand{\max}{\mathrm{max}}
\renewcommand{\over}{\mathrm{o}}
\newcommand{\under}{\mathrm{u}}
\newcommand{\ra}{R}
\newcommand{\mub}{\mu_\mathrm{B}}
\newcommand{\omegap}{\omega_\mathrm{p}}
\newcommand{\co}{c_\mathrm{o}}
\newcommand{\cu}{c_\mathrm{u}}

\usepackage[draft,todonotes={textsize=scriptsize}]{changes} 
\setuptodonotes{inline} 

\usepackage{colortbl}

\definecolor{mustardBody}{RGB}{255, 252, 240}   
\definecolor{mustardHead}{RGB}{245, 230, 180}   

\definecolor{blueABody}{RGB}{245, 250, 255}     
\definecolor{blueAHead}{RGB}{200, 225, 245}     
\definecolor{blueBBody}{RGB}{235, 242, 252}     
\definecolor{blueBHead}{RGB}{180, 210, 240}     
\definecolor{blueCHead}{RGB}{135, 180, 225}     

\definecolor{redABody}{RGB}{255, 248, 245}      
\definecolor{redAHead}{RGB}{255, 215, 205}      
\definecolor{redBBody}{RGB}{255, 240, 235}      
\definecolor{redBHead}{RGB}{255, 190, 180}      
\definecolor{redCHead}{RGB}{255, 140, 125}

\definecolor{grayABody}{RGB}{249, 249, 249}   
\definecolor{grayAHead}{RGB}{222, 222, 222}   

\definecolor{grayBBody}{RGB}{243, 243, 243}   
\definecolor{grayBHead}{RGB}{203, 203, 203}   
\definecolor{grayCHead}{RGB}{176, 176, 176}   

\newcommand{\cpad}{\hspace{8pt}}

\begin{document}
\title{Nonequilibrium Kramers Turnover in a Kerr Parametric Oscillator}
\author{Daniel K. J. Boneß}
	\affiliation{Department of Physics, University of Konstanz, 78464 Konstanz, Germany}

\author{Gabriel Margiani}
    \affiliation{Laboratory for Solid State Physics, ETH Zürich, 8093 Zürich, Switzerland}
	\affiliation{Quantum Center, ETH Zürich, 8093 Zürich, Switzerland}

\author{Wolfgang Belzig}
	\affiliation{Department of Physics, University of Konstanz, 78464 Konstanz, Germany}
    
    \author{Alexander Eichler}
    \affiliation{Laboratory for Solid State Physics, ETH Zürich, 8093 Zürich, Switzerland}
	\affiliation{Quantum Center, ETH Zürich, 8093 Zürich, Switzerland}
	
    \author{Oded Zilberberg}
	\affiliation{Department of Physics, University of Konstanz, 78464 Konstanz, Germany}
	
\begin{abstract}
Activation processes govern noise-induced switching between long-lived states. In an equilibrium double well, the thermally activated switching rate exhibits a prefactor with a nonmonotonic dependence on environmental coupling, a foundational crossover known as Kramers turnover. Here, we demonstrate a Kramers turnover analogue in a Kerr parametric oscillator, a driven-dissipative nonlinear system featuring two stable phase states.  First, we analytically establish turnover physics in this out-of-equilibrium setting. There, the strong physical correlation between the activation barrier and intrinsic damping fundamentally obscures the underlying turnover physics. To overcome this limitation, we rescale the rotating-frame dynamics and introduce a tunable effective friction controlled entirely by the parametric drive. This rescaling comes at the cost of a concurrent rescaling of the effective temperature. Exploiting this simultaneous scaling, we leverage the effective temperature to extract the turnover directly from temperature-dependent observations. Subsequently, measuring noise-induced phase slips in a micro-electromechanical device, we observe a distinct crossover in the prefactor's temperature dependence. Our results unambiguously isolate the out-of-equilibrium turnover regime and highlight that the competition between dissipation and fluctuations profoundly shapes activation dynamics also  beyond equilibrium.
\end{abstract}

\maketitle

\section{Introduction}
Activation processes, ubiquitous in nature, are stochastic transitions between long-lived states induced by environmental noise. The rates of these switching events follow the phenomenological Arrhenius law: the dynamics exhibit a dominant exponential dependence on the ratio of the activation barrier to the noise strength. This simple exponential scaling is often sufficient for modeling macroscopic phenomena ranging from chemical kinetics to climate shifts~\cite{hanggiReactionrateTheoryFifty1990}. Yet, it obscures a more subtle but fundamental dependence on the environmental coupling (damping rate), $\gamma$. The foundational understanding of this effect dates back to H.A. Kramers, who revealed that the preexponential factor of the switching rate exhibits a nonmonotonic scaling with $\gamma$~\cite{kramersBrownianMotionField1940, melnikovKramersTurnover1991}. Specifically, the prefactor is proportional to $\gamma$ in the underdamped regime and scales as $1/\gamma$ in the overdamped limit, see Fig.~\ref{fig:fig1}. This phenomenon, known as Kramers turnover, captures the universal competition between dissipation, which relaxes the system toward local minima, and fluctuations, which drive it over energy barriers. 

The delicate interplay encapsulated by this prefactor scaling remains at the forefront of contemporary physics. Indeed, in spin-boson models~\cite{leggettDynamicsDissipativeTwostate1987} and open quantum systems within the noisy intermediate-scale quantum (NISQ) era, the environmental coupling effectively acts as a continuous measurement apparatus~\cite{wiseman2009quantum, preskill2018quantum}. Consequently, the resulting competition between unitary dynamics and dissipative backaction drives measurement-induced phase transitions~\cite{skinner2019prx, fisher2023annu}. Notably, the saturation of coherence in these monitored systems can be traced back to a cascade of underdamped-to-overdamped transitions in the Liouvillian eigenmodes~\cite{carisch2023prr, carisch2024pra}. This highlights that the core physics of Kramers turnover, specifically the crossover between damping regimes governing the prefactor, continues to shape modern quantum many-body dynamics.

Although Kramers original formulation considered equilibrium systems, noise-induced switching is equally prevalent in driven-dissipative architectures~\cite{dykmanTheoryFluctuationalTransitions1979, dmitrievActivatedTunnelingTransitions1986, marthalerSwitchingRatesParametric2006,chanActivationBarrierScaling2007,chanPathsFluctuationInduced2008,inagaki2016coherent, davidovikj2019high, bohm2019understanding,grimmStabilizationOperationKerrcat2020,margianiExtractingLifetimeSynthetic2022, ng2022phaseflip, hajr2024high, cabral2024roadmap}. A paradigmatic example is the Kerr parametric oscillator (KPO), a nonlinear system subjected to a parametric drive~\cite{dykmanFluctuationalPhaseflipTransitions1998, lifshitzResponseParametricallyDriven2003,paparielloUltrasensitiveHystereticForce2016,leuchParametricSymmetryBreaking2016,gautier2022combined,eichlerClassicalQuantumParametric2023, hajr2024high, cabral2024roadmap}. There, in the appropriate parameter regime, the interplay of driving, nonlinearity, and the damping rate $\gamma$ gives rise to two distinct, stable states of forced vibrations, see Fig.~\ref{fig:fig1}. In the rotating frame, these states manifest as a degenerate double-well potential~\cite{dykmanFluctuationalPhaseflipTransitions1998,marthalerSwitchingRatesParametric2006,margianiThreeStronglyCoupled2025}. Similar to a particle in a static well, stochastic fluctuations, whether thermal or quantum in origin, induce phase-slip transitions between these two dynamic attractors~\cite{lapidusStochasticPhaseSwitching1999,chanPathsFluctuationInduced2008,dykmanFluctuationalPhaseflipTransitions1998,marthalerSwitchingRatesParametric2006,chanActivationBarrierScaling2007,margianiExtractingLifetimeSynthetic2022, hajr2024high, bohm2019understanding, ng2022phaseflip,frattiniObservationPairwiseLevel2024}. 

Despite extensive studies of phase slips in KPOs, identifying a Kramers turnover in such strongly nonequilibrium systems remains a major theoretical and experimental challenge. Similar to equilibrium settings, the excitation dynamics of a KPO can be decomposed into a conservative contribution, associated with an effective nonequilibrium potential landscape (the quasipotential), and a nonconservative dissipative contribution arising from environmental coupling. As a result, the dynamics exhibit both underdamped and overdamped characteristics~\cite{dykmanFluctuationalPhaseflipTransitions1998,heugelRoleFluctuationsQuantum2023}. However, in nonequilibrium systems the phase-space positions of the attractors depend explicitly on the dissipative force, quantified by the intrinsic damping rate $\gamma$. Consequently, the activation barrier itself acquires a highly nonlinear dependence on $\gamma$. This strong correlation dictates a complex dependence already in the switching rate's exponential dependence. Depending whether the system is in the under or overdamped regime, different power law dependencies and critical exponents have been observed \cite{chanActivationBarrierScaling2007}. Historically, this was presumed to completely mask any subtle prefactor scaling. Correspondingly, a theoretical derivation establishing whether a fundamental $\gamma$ to $1/\gamma$ crossover even exists in the prefactor has remained absent from the literature. Furthermore, measurements of the macroscopic switching rate alone are insufficient to unambiguously identify the turnover or to meaningfully compare dynamics across damping regimes.

Here, we demonstrate a Kramers turnover analogue in the nonequilibrium dynamics of a  KPO. We overcome the aforementioned theoretical and experimental challenges by mapping the dynamics in a way such that the conservative part remains constant, while changing the effective dissipation rate. Furthermore, we derive the analytical form of the underdamped switching rate. This reveals a prefactor proportional to $\gamma$, which contrasts sharply with the $1/\gamma$ dependence established for the overdamped regime~\cite{dykmanFluctuationalPhaseflipTransitions1998}. To circumvent the experimental inability to arbitrarily tune the physical damping rate, we introduce effective friction and temperature parameters that are controlled  by the parametric drive amplitude and frequency. Subsequently, by measuring noise-induced switching rates in a micro-electromechanical device, we sweep continuously between the effective over- and underdamped regimes while keeping the intrinsic physical damping rate $\gamma$ fixed. This methodology reveals a clear crossover in the temperature dependence of the prefactor, isolating the turnover. Our results, thus, extend the competition between dissipation and fluctuations far beyond equilibrium systems, providing a unified framework for understanding activation in out-of-equilibrium environments.

The remainder of this manuscript is organized as follows. In Sec.~\ref{sec:models}, we introduce the theoretical models governing the relevant activation dynamics. First, we review the foundational Kramers turnover in a static double-well potential to establish the equilibrium baseline. Subsequently, we detail the strongly nonequilibrium dynamics of the  KPO. There, we derive the analytical form of the underdamped switching rate and outline the theoretical obstacles to isolating the prefactor scaling. Moving on, Sec.~\ref{sec:results} presents our primary experimental and numerical results. First, we formulate a rescaling of the rotating-frame dynamics to introduce a tunable effective friction. We then present measurements performed on a micro-electromechanical resonator. By analyzing temperature-dependent phase-slip statistics, we unambiguously isolate the out-of-equilibrium turnover regime. Finally, Sec.~\ref{sec:conclusion} summarizes our conclusions and discusses broader physical implications. Furthermore, extensive details regarding (i) static system scaling (Appendix~\ref{app_kramers_static}), (ii) numerical simulations (Appendix~\ref{app_numerical_simulation}), (iii) analytical rate derivations (Appendix~\ref{app_switching_KPO}), and (iv) experimental calibration (Appendix~\ref{app_calibration}) are relegated to the Appendices. Throughout the manuscript, we introduce various symbols and notations that are summarized in  Table~\ref{tab:variables}.

\section{Models}\label{sec:models}
\begin{figure}[t!]
	\centering
	\includegraphics[width=1\linewidth]{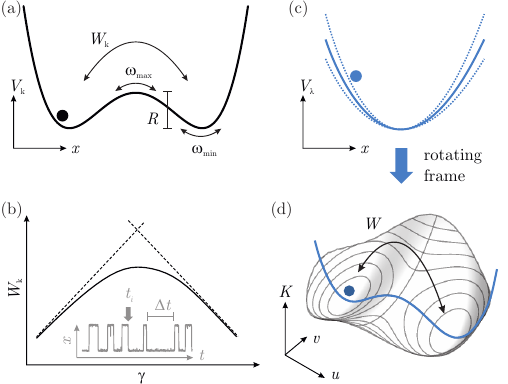}
	\caption{\textit{Activated switching in equilibrium and out of equilibrium}. (a) Particle in a double-well potential with barrier $R$ and local extrema frequencies $\omega_{\rm min}$ and $\omega_{\rm max}$ [cf.~Eqs.~\eqref{eq:general_eom} and~\eqref{eq:double_well_pot}], subject to interwell switching at rate $W_{\rm K}$ [cf.~Eq.~\eqref{eq:general_rates}]. (b) Switching rate $W_{\rm K}$ versus damping rate $\gamma$, including over- and underdamped limits (dashed lines) from Eqs.~\eqref{eq_static_over} and~\eqref{eq_static_under}. Inset: Particle position versus time exhibiting stochastic switching events, e.g., at time $t_i$ (arrow), separated by dwell times, $\Delta t$, in a given state. From such data, we extract the activation rate~\cite{margianiExtractingLifetimeSynthetic2022} (see App.~\ref{app_numerical_simulation}). (c) Schematic of a  KPO featuring time-varying potential stiffness [cf.~Eqs.~\eqref{eq:general_eom} and~\eqref{eq:KPO_pot}]. (d) Rotating-wave approximation Hamiltonian of the oscillator in rotating phase space $(u,v)$ [cf.~Eq.~\eqref{eq:RWA_pot}], where phase-slips occur as activation with rate $W$.}
	\label{fig:fig1}
\end{figure}
We consider the dynamics of a massive particle moving under the influence of a time-dependent potential $V(x,t)$ while coupled to an environment,
\begin{align}
\label{eq:general_eom}
    \ddot{x} + \partial_x V(x,t) + \gamma \dot{x} = \xi(t)\,,
\end{align}
where $x$ is a spatial coordinate and dots signify derivation with respect to time $t$. In this work, we will compare two seemingly very different cases: on the one hand, we will revisit the equilibrium setting of a static double-well potential [see Fig.~\ref{fig:fig1}(a)],
\begin{align}
\label{eq:double_well_pot}
    V(x,t)\rightarrow V_{\mathrm{k}}(x)=a x^4-bx^2\,,
\end{align}
where $a,b>0$. On the other hand, we will study the time-dependent out-of-equilibrium case of a KPO~\cite{eichlerClassicalQuantumParametric2023} [see Fig.~\ref{fig:fig1}(b)],
\begin{multline}
    \label{eq:KPO_pot}
    V(x,t)\rightarrow \\ V_\lambda(x,t) = \omega_0^2 [1-\lambda \cos(2\omegap t)] x^2/2 + \alpha x^4/4\,,
\end{multline}
where $\omega_0$ is the angular eigenfrequency and $\alpha$ sets the scale of the nonlinear potential. Furthermore, $\lambda$ and $2\omegap$ denote the depth and angular frequency of the parametric drive, respectively. In both cases, we are interested in the effect of coupling our system to the environment, which gives rise to two distinct forces in Eq.~\eqref{eq:general_eom}: a deterministic friction term $\gamma \dot{x}$ and a stochastic force $\xi(t)$. Throughout, we assume Gaussian white noise characterized by $\langle \xi(t)\xi(t')\rangle = 2D\,\delta(t-t')$. In equilibrium settings, damping and noise share a common physical origin and are therefore linked by the fluctuation–dissipation theorem, yielding $D=\gamma \kb T/m$, where $\kb$ is Boltzmann’s constant, $T$ the environmental temperature, and $m$ the particle's effective mass, which we set to unity in the following.

\subsection{Kramers turnover}
We begin by revisiting the well-studied equilibrium case of the static double-well~\cite{kramersBrownianMotionField1940}, cf.~Eq.~\eqref{eq:double_well_pot}. Here, dissipation renders the two potential minima dynamically stable, turning them into attractors of the motion, see Fig.~\ref{fig:fig1}(a). At the same time, the stochastic force $\xi(t)$ induces random perturbations that drive the system away from these attractors. In the weak-noise regime, the system remains confined to a single potential well for the vast majority of the time, and transitions between wells occur only rarely, see inset of Fig.~\ref{fig:fig1}(b). Such switching events require a sequence of fluctuations (stochastic force kicks) sufficiently strong to overcome the barrier separating the wells and are therefore exponentially unlikely. Among all possible fluctuation realizations leading to escape, the dominant contribution arises from a single trajectory, commonly referred to as the \textit{least improbable escape path} (LIEP). In the weak-noise limit, fluctuation probabilities obey a large-deviation principle, where each noise realization is weighted by an action functional. The LIEP minimizes this action, and the corresponding minimum defines the effective activation energy $\ra$. Escape events are dominated by this optimal trajectory and the rate of switching events depends exponentially on the ratio of the effective activation energy $\ra$ to the thermal energy $\kb T$~\cite{freidlinRandomPerturbationsDynamical}
\begin{align}
	\label{eq:general_rates}
	W = C \,\exp[-\ra/\kb T]\,,
\end{align}
where $C$ is a prefactor that depends smoothly on the system parameters. For constant $C$, Eq.~\eqref{eq:general_rates} constitutes the Arrhenius law~\cite{arrhenius_1889_1749766}. 

\begin{figure}[t!]
	\centering
	\includegraphics[width=1\linewidth]{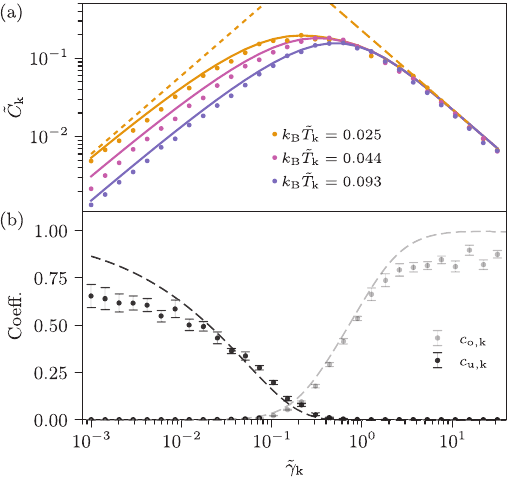}
	\caption{\textit{Prefactor for the static double-well potential}. (a) Rescaled prefactor $\tilde C_\mathrm{k}$ [cf.~Eq.~\eqref{eq_equilibrium_rate} and Table~\ref{tab:variables}] of the double-well switching rate as a function of rescaled damping $\tilde{\gamma}_{\rm k}$ (see App.~\ref{app_kramers_static}), extracted from numerical simulations (crosses, see App.~\ref{app_numerical_simulation}) and compared with (i) the general expression of Eq.~\eqref{eq_melnikov} (solid lines) for different effective temperatures $\tilde{T}_{\rm k}$; and (ii) the overdamped and underdamped limits [Eqs.~\eqref{eq_static_over} and~\eqref{eq_static_under}; dashed and dotted lines, respectively] evaluated at $\kb \tilde{T}_\mathrm{k} = 0.025$ [cf.~Fig.~\ref{fig:fig1}(b)]. (b) Coefficients $c_\mathrm{o,k}$ and $c_\mathrm{u,k}$ extracted via fits of the rescaled prefactor to Eq.~\eqref{eq_fit_static} as a function of temperature (see App.~\ref{app:Kramers_interpolate}), comparing numerical simulations (crosses) with fits to the analytical expression of Eq.~\eqref{eq_equilibrium_rate} (dashed line). Rescaled quantities marked with a tilde are defined in App.~\ref{app_kramers_static}.}
	\label{fig:fig2}
\end{figure}

For the static double-well potential~\eqref{eq:double_well_pot}, the activation energy is given by the height of the potential barrier, $\ra_\mathrm{k} = V_\mathrm{k}(x_\mathrm{max}) - V_\mathrm{k}(x_\mathrm{min})$~\cite{kramersBrownianMotionField1940}, where $x_{\mathrm{min/max}}$ are the coordinates of the potential extrema. Hereafter, we use the subscript $\mathrm{k}$ for quantities related to the static potential, cf.~Table~\ref{tab:variables}. Crucially, the dependence of the switching rate $W_\mathrm{k}$ on $\gamma$ is contained  in the prefactor~\cite{melnikovKramersProblemFifty1991}
\begin{align}
	C_\mathrm{k} = \frac{\omega_\min}{2\pi} \left[\sqrt{1+ \frac{\gamma^2}{4\omega_\max^2} } - \frac{\gamma}{2 \omega_\mathrm{max}}\right] \frac{A(\gamma S/T)^2}{A(2 \gamma S/T)}\,,
\label{eq_equilibrium_rate}
\end{align}
where $\omega_{\mathrm{min/max}} = |\partial_x^2 V_\mathrm{k}(x_{\mathrm{min/max}})|^{1/2}$ denote the vibration frequencies close to the potential's extrema, and $S$ describes the action per oscillation of a particle near the potential maximum. The explicit form of the function $A(x)$ is given in Appendix~\ref{app_kramers_static}, leading to a smooth interpolation between two limits in Eq.~\eqref{eq_equilibrium_rate}: regimes where $\gamma$ is either   large or small compared with $T/S$ and $\omega_{\mathrm{min}}$~\footnote{This scaling arises because $\omega_\mathrm{min} = \sqrt{2} \omega_\mathrm{max}$}. For small $\gamma/\omega_{\rm min}$, the deterministic motion around the minima is underdamped and $C_\mathrm{k} \propto \gamma$. Conversely, for large $\gamma/\omega_{\rm min}$, the motion is overdamped and $C_\mathrm{k} \propto 1/\gamma$. In both asymptotic limits, the switching rate $W_\mathrm{k}$ is heavily suppressed. This results in a maximum of $W_\mathrm{k}$ at intermediate damping strengths, see Figs.~\ref{fig:fig1}(b) and~\ref{fig:fig2}(a). This non-monotonic crossover of the prefactor is known as Kramers turnover~\cite{kramersBrownianMotionField1940}. Historically, this turnover proved difficult to isolate because tuning $\gamma$ over a wide, continuous range alters the system dynamics and presents a severe experimental limitation. Notwithstanding, recent advances have successfully confirmed these predictions~\cite{rondinDirectMeasurementKramers2017}, inspiring also connections to the physics of measurement-induced phase transitions~\cite{carisch2023prr,carisch2024pra}.

Conventionally, Kramers turnover is observed by sweeping the damping rate across a wide parameter space at a fixed temperature. Here, we introduce an alternative methodology to identify the turnover directly through the distinct temperature dependence of the prefactor. Specifically, the prefactor exhibits a pronounced $1/T$ scaling in the underdamped limit, whereas this temperature dependence vanishes  in the overdamped regime, see Fig.~\ref{fig:fig2}(a). We exploit this qualitative distinction to quantify the physical crossover: we fit the extracted prefactor as a function of temperature to a direct superposition of its asymptotic limiting forms, see Appendix~\ref{app:Kramers_interpolate}. To this end, we introduce the phenomenological fitted coefficients $c_\mathrm{u,k}$ and $c_\mathrm{o,k}$ that measure the relative weights of the temperature-dependent (underdamped) and temperature-independent (overdamped) contributions, respectively, see Eq.~\eqref{eq_fit_static} in Appendix~\ref{app_kramers_static}. Ultimately, the turnover is captured  by the values of these coefficients. In the underdamped regime, we find $c_\mathrm{u,k}\approx 1$ and $c_\mathrm{o,k}\approx 0$, whereas the opposite limits are reached in the overdamped regime, see Fig.~\ref{fig:fig2}(b). Note that in the intermediate region, both limiting models predict significantly higher rates than Eq.~\eqref{eq_equilibrium_rate} (cf.~Fig.~\ref{fig:fig1}(b)). As a result, the sum of both coefficients is $<1$. Despite a systematic underestimation of the switching rates in the underdamped regime [see Fig.~\ref{fig:fig2}(a)], the turnover remains clearly identifiable from both the prefactors themselves and the fitted coefficients shown in Fig.~\ref{fig:fig2}(b).

The bare prefactor $C_\mathrm{k}$ and the activation energy $\ra_\mathrm{k}$ depend explicitly on the double-well parameters $a$ and $b$. To generalize our findings, we map the system onto a parameter-free effective potential. This transformation yields dimensionless rates characterized by a rescaled prefactor $\tilde{C}_\mathrm{k}$ and a rescaled activation energy $\tilde{\ra}_\mathrm{k}$. Correspondingly, the modified dynamics are governed  by a rescaled temperature $\tilde{T}_\mathrm{k}$ and an effective dissipation rate $\tilde{\gamma}_\mathrm{k}$. This raises an interesting avenue towards Kramers physics: we can systematically tune the effective dissipation by varying the potential landscape via $a$ and $b$, even while the intrinsic physical damping $\gamma$ remains    constant, see Appendix~\ref{app_kramers_static} and Table~\ref{tab:variables}. Such an effective parameter mapping proves highly advantageous for analyzing the driven-dissipative dynamics of the KPO. Moving forward, we establish a similar nonmonotonic crossover in the prefactor governing the switching between the forced oscillation states of Eq.~\eqref{eq:KPO_pot}. To achieve this, we first detail the complex interplay between the nonlinear time-dependent potential $V_{\lambda}$, its dynamically shifted attractors, and $\gamma$.

\subsection{Activation in a KPO}
We now turn to the nonequilibrium system described in Eq.~\eqref{eq:KPO_pot}~\cite{dykmanFluctuationalPhaseflipTransitions1998, eichlerClassicalQuantumParametric2023}. Here, the parametric modulation with depth $\lambda$ and frequency $2\omegap$ excites vibrations in the system by periodically changing the curvature around the minimum, see Fig.~\ref{fig:fig1}(c). Without loss of generality, we assume negative nonlinearity $\alpha<0$. Under typical assumptions where $\lambda, \gamma/\omega_0$, and $-\alpha x^2 /\omega_0^2$ are assumed to be small, the system is well approximated in a rotating frame: we introduce the slow-flow coordinates $u,v$ defined by $x = u \cos(\omegap t) - v \sin(\omegap t)$, which under the rotating wave approximation (RWA) evolve as~\cite{eichlerClassicalQuantumParametric2023}
\begin{align}
	\begin{split}
		\dot{u} &= \frac{\partial K(v,u)}{\partial v} - \frac{\gamma}{2} u + \frac{1}{\sqrt{2} \omegap } \xi_u(t) \,,\\
		\dot{v} &= -\frac{\partial K(v,u)}{\partial u} - \frac{\gamma}{2} v + \frac{1}{\sqrt{2} \omegap } \xi_v(t)\,. 
		\label{eq:slow_flow}
	\end{split}
\end{align}
The noise processes $\xi_v$ and $\xi_u$ are approximately uncorrelated, with~\cite{eichlerClassicalQuantumParametric2023}
\begin{align}
\label{eq_xi_KPO}
{\langle \xi_{i}(t) \xi_{j}(t') \rangle \approx 2 D \delta_{ij} \delta (t-t')}.
\end{align}
The RWA Hamiltonian (also dubbed quasienergy potential) is given by
\begin{multline}
K(u,v) = -\frac{3\alpha}{32 \omegap}(u^2+v^2)^2 \\+ \frac{\lambda \omega_0^2}{8 \omegap} [u^2 (1-\mu) - v^2(1+\mu)]\,,
\label{eq:RWA_pot}
\end{multline}
where we introduced the scaled detuning 
\begin{align}
    \label{eq:mu}
    \mu = 2 (\omega_0^2 - \omegap^2) /\lambda \omega_0^2.
\end{align}
In the regime where $-1 < \mu < 1$, the RWA Hamiltonian~\eqref{eq:RWA_pot}, $K$, exhibits two minima separated by a saddle point, see Fig.~\ref{fig:fig1}(d). Translating back to the laboratory frame, these minima represent steady-state oscillations with identical amplitudes but a relative $\pi$ phase shift, commonly termed ``phase states''.  Crucially, due to the competition between parametric driving and dissipation, the stable attractors of the full slow-flow dynamics~\eqref{eq:slow_flow} are shifted away from the minima of $K$, see Appendix~\ref{app_rescaling} and Ref.~\cite{dumontEnergyLandscapeFlow2024,bachtoldMesoscopicPhysicsNanomechanical2022}. As a consequence, the parameter range supporting two stable attractors is reduced and given by the condition $\mub^{(-)} < \mu < \mub^{(+)}$, where
\begin{align}
	\mub^{(\mp)} = \mp \sqrt{1-4\frac{\omegap^2 \gamma^2}{\lambda^2 \omega_0^4}}\,.
\end{align}
In the following, we focus on the bistable regime, where the KPO displays the two stable phase states. Similar to the particle in the double-well potential, fluctuations can induce transitions between the two states, which correspond to phase-flips in the response of the KPO.

We investigate the rate of thermally induced phase-flip transitions in the weak-noise limit. Analytical expressions for the activation rate can be obtained in two limiting cases, corresponding to underdamped and overdamped dynamics in the rotating frame~\cite{dykmanFluctuationalPhaseflipTransitions1998}. In the driven system, these regimes are distinguished by comparing the effective oscillation frequency near a stable attractor with the dissipation rate~$\gamma$. The effective frequency is set by the curvature of the RWA Hamiltonian $K(v,u)$ at its minima and is given by
\begin{align}
\label{eq_omega_min_KPO}
    \Omega_{\mathrm{min}} = \frac{\lambda \omega_0^2}{2\omegap}\sqrt{1+\mu}\,.
\end{align}
For sufficiently large drive amplitudes $\lambda$, the system resides in an underdamped regime: $\Omega_{\mathrm{min}} \gg \gamma$. In stark contrast to static systems, the KPO fundamentally lacks a bistable phase where $\Omega_{\mathrm{min}} \ll \gamma$. Consequently, a deeply overdamped limit cannot be physically realized without destroying the bistability. Notwithstanding, the dynamics become effectively overdamped near the critical bifurcation threshold: the system satisfies $\Omega_{\mathrm{min}} \sim \gamma$ within the narrow parameter window $0 < \mu - \mub^{(-)} \ll 1$. For further mathematical details regarding this effective regime, we refer the reader to Ref.~\cite{dykmanFluctuationalPhaseflipTransitions1998} and Appendix~\ref{app_rates_over}.

Analytical expressions for the activation energy and the corresponding prefactor in the overdamped regime are firmly established in the literature~\cite{dykmanFluctuationalPhaseflipTransitions1998}. Conversely, historical treatments of the underdamped regime successfully identified the activation energy but omitted the associated prefactor entirely~\cite{dykmanFluctuationalPhaseflipTransitions1998}. We investigate this underdamped regime in Appendix~\ref{app_rates_under}. There, extending beyond the known exponential scaling, we analytically derive this previously absent prefactor. Consolidating these established and novel findings yields the respective transition rates:
\begin{align}
	W_\mathrm{u} &\approx \frac{\gamma}{\kb T} \frac{4 \sqrt{1+\mu} I_s(\mu) \lambda \omega_0^2 \omegap^2 }{ (2+\mu) 3 (-\alpha) } \exp\left[-\frac{\ra_\mathrm{u}}{\kb T}\right],\label{eq_rate_under}\\
	W_\mathrm{o} &\approx \frac{1}{\gamma}\frac{(\mu-\mub^{(-)}) |\mub^{(-)}| \lambda^2 \omega_0^4}{8 \sqrt{2} \pi  \omegap^2} \exp\left[-\frac{\ra_\mathrm{o}}{\kb T}\right]\,,\label{eq_rate_over}
\end{align}
for the underdamped and overdamped regime, respectively. In the overdamped regime the activation energy is given by
\begin{align}
    \ra_\mathrm{o} &= \frac{|\mub^{(-)}| (\mu - \mub^{(-)})^2 \lambda \omega_0^2 \omegap^2}{ -6 \alpha}.
\end{align}
The expressions for $\ra_\mathrm{u}$ and $I_s$ appear in Appendix~\ref{app_rates_under}.

One of the central results of this work is the analytical expression for the underdamped transition rate prefactor in Eq.~\eqref{eq_rate_under}. Crucially, we establish that this term scales linearly with the damping rate $\gamma$ and exhibits an explicit inverse dependence on temperature, perfectly mirroring the underdamped scaling in the static double-well case, cf.~Eq.~\eqref{eq_static_under}. This behavior contrasts starkly with the overdamped regime: on the one hand, the prefactor remains    independent of temperature, which is  consistent with the static equilibrium system, see Fig.~\ref{fig:fig2} and Eq.~\eqref{eq_static_over}. On the other hand, the analytical scaling with $\gamma$ in this nonequilibrium system is highly intricate: the bifurcation parameter $\mub^{(-)}$ itself evolves dynamically with the damping~\cite{dykmanFluctuationalPhaseflipTransitions1998, dykmanScalingActivatedEscape2005}. Notwithstanding, the resulting overdamped prefactor decreases monotonically with increasing $\gamma$, scaling approximately as $\sim 1/\gamma$. Combining the linearly increasing underdamped limit with this monotonically decreasing overdamped limit guarantees that the overall prefactor peaks at intermediate environmental coupling. Ultimately, this mathematical competition dictates that the driven-dissipative KPO must exhibit a Kramers turnover analogue. To confirm this prediction, several physical challenges must first be addressed:
\begin{enumerate}
\item Similar to equilibrium settings, tuning $\gamma$ over a wide range presents a severe experimental limitation. Instead, we maintain a fixed physical $\gamma$ while  varying the parametric drive parameters $\lambda$ and $\omegap$. This explicit variation alters the conservative RWA Hamiltonian $K$. Turning a liability into an asset, we introduce rescaled potential parameters to map the dynamics such that the effective damping is swept instead without modifying the shape of the underlying potential landscape. This methodology enables a fair comparison of the switching rates and allows us to sweep continuously between the effective over- and underdamped limits. 
\item Unlike the equilibrium case of Eq.~\eqref{eq_equilibrium_rate}, the nonequilibrium activation energy also displays a highly nonlinear dependence on $\gamma$ in the overdamped regime. This nonlinearity arises directly from the dissipation-induced shifting of the attractors in phase space~\cite{dykmanFluctuationalPhaseflipTransitions1998, dykmanScalingActivatedEscape2005}. Because the exponential argument of $W_{\mathrm{o}}$ scales explicitly with $\gamma$, it fundamentally obscures the comparatively weak prefactor scaling. Hence, identifying the turnover requires isolating the prefactors via their distinct temperature dependencies at fixed effective friction. 
\item Finally, because the experimentally accessible parameter space remains    bounded, we supplement our physical demonstration with numerical simulations.
\end{enumerate}

\begin{figure*}[t!]
	\centering
	\includegraphics[width=1\linewidth]{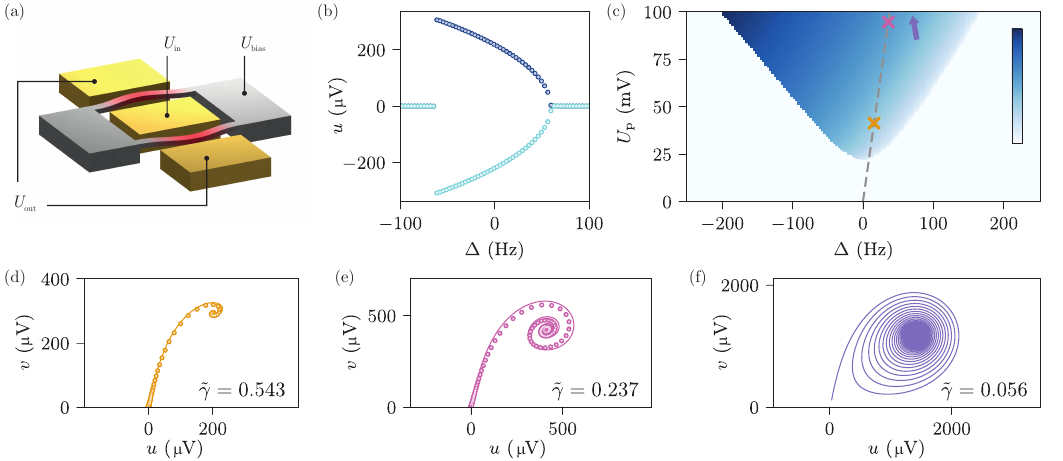}
	\caption{\textit{Characterization of the KPO}. (a) Schematic of the experimental device with a bias voltage $U_{\mathrm{bias}}$ application, capacitive actuation $U_{\mathrm{in}}$, and readout of the displacement-generated voltage signal $U_{\mathrm{out}} = u \cos(\omega_\mathrm{p} t) - v \sin(\omega_\mathrm{p} t)$. (b) Measured in-phase quadrature $u$ of the parametrically driven resonator versus detuning $\Delta = (\omegap-\omega_0)/2\pi$ at fixed drive amplitude $U_\mathrm{p} = 0.0388\,\mathrm{V}$, featuring two distinct frequency sweeps (bright and dark symbols) to reveal the amplitude-degenerate phase states. (c) Forced vibration amplitude, $A\equiv \sqrt{u^2+v^2}$, versus $\Delta$ and $U_\mathrm{p}$, including the constant $\mu = -0.2$ contour (gray dashed line). (d)--(e) Ringdown measurements along the constant $\mu = -0.2$ contour for varying effective damping rates $\tilde{\gamma}$ [crosses in panel~(c)], comparing experimental data (circles) with noiseless numerical simulations of Eqs.~\eqref{eq:slow_flow} (solid lines, cf.~App.~\ref{app_numerical_simulation}). (f) Numerical ringdown simulation for a deeply underdamped regime [direction purple arrow in panel~(c)]. }
	\label{fig:fig3}
\end{figure*}

\section{Results}\label{sec:results}
Having established the theoretical framework for a Kramers turnover analogue in the KPO, we confirm this prediction via experimental observation and numerical simulation. We first recast the model using  rescaled parameters, and then turn to the protocol for the observation of the effect.
\subsection{The secret to observing Kramers turnover in a KPO}
Varying the experimental drive parameters $\lambda$ and $\omegap$ fundamentally alters the physical RWA Hamiltonian $K$ [cf.~Eq.~\eqref{eq:RWA_pot}]. Specifically, these physical variations dynamically modify both the barrier height, $R$ and the local minimum frequency $\Omega_{\min}$, profoundly impacting the raw activation rates~\eqref{app_rates_under} and~\eqref{app_rates_over}. To disentangle these effects from the KPO's Kramers turnover, we  recast the model using rescaled parameters, i.e., we rescale time and the phase-space quadratures, see Appendix~\ref{app_rescaling}. This transformation maps the varying physical landscape onto a dimensionless effective potential governed by a single parameter: the rescaled detuning $\mu$~\cite{dykmanFluctuationalPhaseflipTransitions1998}, cf.~Eq.~\eqref{eq:mu}. This formal procedure shifts the complex parameter dependence into an effective friction coefficient and an effective temperature,
\begin{align}
\label{eq:gamma_eff}
	\gamma &\rightarrow \tilde \gamma =  2\omegap \gamma/\lambda \omega_0^2,\\
	\label{eq:T_eff}
	T &\rightarrow \tilde T = -3 \alpha T/2 m \lambda \omega_0^2\omegap^2.
\end{align}

This rescaling provides a highly advantageous experimental architecture: we can isolate specific trajectories in the $(\omegap,\lambda)$ parameter space where the effective detuning $\mu$ is held constant. Along these targeted trajectories, the rescaled potential remains structurally identical, whereas the effective damping $\tilde \gamma$ is independently tuned. Note that we can relate this scaling architecture to an analogous construction in the static equilibrium system, see Appendix~\ref{app_kramers_static}.

From the rescaled parameters, we define the dimensionless switching rates $\tilde W$ by normalizing the physical rates against the prefactor $\lambda \omega_0^2 / 4 \omegap$. Consolidating these definitions, Eqs.~\eqref{eq_rate_under} and~\eqref{eq_rate_over} take the explicit dimensionless form
\begin{align}
	\tilde W_\under &= \frac{\tilde \gamma}{\kb \tilde T}\frac{2 \sqrt{1+\mu} I_s(\mu) }{(2+\mu)}  \exp \left[-\frac{\tilde \ra_\under}{\kb \tilde T}\right]\,, \label{eq_rate__dimless_under}\\
	\tilde W_\over &= \frac{1}{\tilde \gamma}\frac{(\mu-\mub^{(-)}) |\mub^{(-)}|}{\sqrt{2}\pi } \exp \left[-\frac{\tilde \ra_\over}{\kb \tilde T}\right]\,. \label{eq_rate__dimless_over}
\end{align}
Hence, our rescaling procedure provides a systematic framework to interpret the highly complex transition rate expressions. While the isolated prefactor  peaks at intermediate coupling, this Kramers-like turnover does not manifest directly as a maximum in the total observable transition rate. This is due to the rescaled activation energy $\tilde \ra_\mathrm{o}$ explicitly depending on the effective damping $\tilde{\gamma}$, and the resulting exponential therefore fundamentally overpowering the comparatively weak crossover of the prefactor. Notwithstanding, this obscured turnover can be successfully revealed by turning the concurrent temperature rescaling into a diagnostic feature. Specifically, the prefactor remains independent of the rescaled temperature $\tilde{T}$ in the overdamped regime, whereas it scales as $1/\tilde{T}$ in the underdamped limit.

\subsection{Experiment}
We set out to extract the activation prefactors and observe the KPO's Kramers turnover both experimentally and through numerical simulations. Together, this provides compelling evidence that a fundamental Kramers turnover manifests in the driven-dissipative activation dynamics of the KPO.

Our experimental device is a micro-electromechanical resonator, see Fig.~\ref{fig:fig3}(a)~\cite{agarwal2008study}. It was shown previously that these devices are faithfully described by the KPO equation of motion~\eqref{eq:slow_flow}~\cite{miller2019phase,dumontEnergyLandscapeFlow2024}. The device comprises a mechanical double-ended tuning fork resonator made from doped silicon. It is encapsulated in mild vacuum (\SI{1e-1}{\milli\bar}) on a chip and only accessible through the electrodes surrounding the resonator. By applying a bias voltage \( U_{\text{bias}} \) to the tuning fork itself, the resonator's eigenfrequency \( \omega_0 \) can be tuned through the electrostatic force between the mechanical element and the electrodes, and a Duffing non-linearity \( \alpha \) is induced due to the nonlinearity of this force. A voltage \( U_{\text{in}} = U_\mathrm{p} \cos(2 \omegap t) \) on the two outer electrodes periodically modulates the resonator's eigenfrequency with a modulation depth \( \lambda \propto U_\mathrm{p} \), thereby enabling parametric driving of the resonator. The mechanical displacement \( x \) is detected as a varying voltage on the center electrode as \( U_{\text{out}} \propto x \) using a lock-in amplifier (Zurich Instruments MFLI). For convenience, we forgo the proportionality coefficient and define \( U_{\text{out}} = x \) in our treatment. 

We perform four distinct experimental protocols with our device: (i)~deterministic detuning sweeps to calibrate the system and map the parameter regime supporting stable phase states, (ii)~deterministic rotating-frame ringdown measurements to visualize the effective damping $\tilde{\gamma}$ and the local angular speed $\tilde \Omega_\mathrm{min}$ near a KPO phase state, (iii)~stochastic switching measurements to accurately capture the phase-slip activation rates, and (iv)~systematic extractions of these rates as a function of both the effective temperature $\tilde{T}$ and the effective damping $\tilde{\gamma}$. Consolidating these comprehensive techniques allows us to isolate the prefactors governing the asymptotic underdamped and overdamped limits. This approach demonstrates the targeted Kramers turnover within the driven-dissipative KPO. 

We first characterize our KPO by sweeping the detuning $\Delta = (\omegap-\omega_0)/2\pi$ from low to high frequencies at a fixed drive amplitudes. An example for $U_\mathrm{p} = 0.0388\,\mathrm{V}$, is shown in Fig.~\ref{fig:fig3}(b). There, around $\Delta\sim -61$~Hz, the measured response amplitude abruptly jumps from zero to $561\,\mu\mathrm{V}$: this discontinuity marks a bifurcation point of the slow-flow dynamics in Eqs.~\eqref{eq:slow_flow}. There, the resonator experiences \textit{parametric resonance} and undergoes a spontaneous time-translation symmetry-breaking phase transition~\cite{heugelClassicalManyBodyTime2019,soriente2021,heugelRoleFluctuationsQuantum2023,eichlerClassicalQuantumParametric2023}, i.e., the system leaves the trivial zero-amplitude state and settles into one of the two equal-probability phase states, see Fig.~\ref{fig:fig1}(d). Moving forward, these driven phase states terminate in a second continuous bifurcation at $\Delta\sim 60$~Hz. By iterating these systematic sweeps for varying $U_\mathrm{p}$, we map the boundaries of the so-called Arnold tongue, i.e., the parameter regime where the system undergoes parametric resonance, see Fig.~\ref{fig:fig3}(c). Fitting the theoretical shape of this stability lobe and its internal response amplitude to our experimental data allows us to extract the fundamental system parameters. Specifically, we obtain $\omega_0/2\pi \approx 1.13~\text{MHz}$, $\gamma \approx 537~\text{Hz}$, and $\alpha \approx -4.4 \times 10^{16}~\text{V}^{-2}\text{s}^{-2}$, see Appendices~\ref{app_rescaling} and~\ref{app_calibration} for more comprehensive calibration details. 
\begin{figure*}[t!]
	\centering
	\includegraphics[width=1\linewidth]{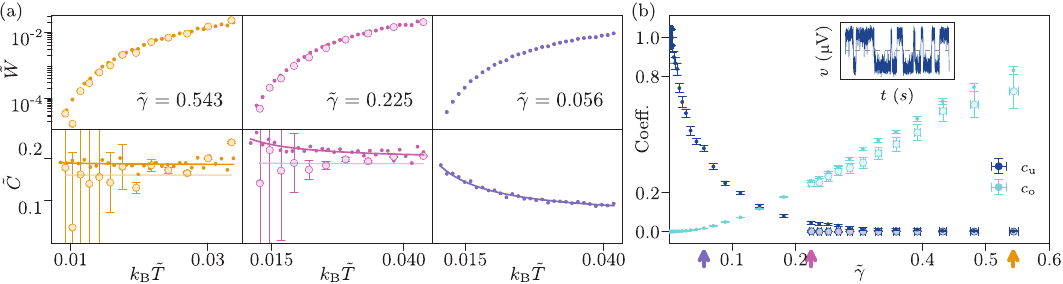}
	\caption{ \textit{Kramers turnover behavior in a KPO}. 
    (a) Scaled switching rate $\tilde{W}$ (top row) as a function of the scaled temperature for different values of $\tilde{\gamma}$, together with the corresponding scaled prefactor $\tilde{C}$ extracted from the data (bottom row) using Eq.~\eqref{eq_prefactor_KPO_scaled}, see Appendix~\ref{app_switching_KPO} for the calculation of $\tilde R$. Transparent and solid lines show fits using Eq.~\eqref{eq_fit_KPO}, with $c_\mathrm{o}$ and $c_\mathrm{u}$ treated as free parameters, for the experimental and simulated data, respectively. 
    (b) Prefactor coefficients $c_\mathrm{o}$ and $c_\mathrm{u}$ that were extracted in panel (a), as a function of $\tilde{\gamma}$. Arrows indicate the values of $\tilde{\gamma}$ corresponding to the datasets shown in panel~(a).
    In both panels large circles indicate experimental measurements, while small dots correspond to simulated data according to App.~\ref{app_numerical_simulation}. Note that no measurement was possible for $\tilde{\gamma}<0.2$.}
	\label{fig:fig4}
\end{figure*}

In a second experiment, we map the transition between over- and underdamped dynamics in the rotating frame by performing ringdown experiments within the driven quasi-potential~\cite{dumontEnergyLandscapeFlow2024,villaTopologicalClassificationDrivendissipative2024}, cf.~Fig.~\ref{fig:fig1} and Eq.~\eqref{eq:RWA_pot} for the quasi-potential and see Figs.~\ref{fig:fig3}(d)-(f) for ringdown examples. For each ringdown, the system is initialized at its equilibrium position in the absence of forcing, $(q,p) = (u,v) = (0,0)$. Once the parametric drive is turned on, the system evolves under the continuous influence of the parametric drive and damping, cf.~Eqs.~\eqref{eq:slow_flow}.
Following these slow-flow dynamics, the oscillator approaches one of the stable attractors along a spiraling trajectory. There, the classification as overdamped or underdamped depends on a dynamical ratio: the rate of radial decay toward the attractor relative to the angular frequency of the spiral. In our rescaled coordinates, these quantities correspond directly to the effective damping $\tilde \gamma$ and the effective angular speed $\tilde \Omega_\mathrm{min} = 2 \sqrt{1+\mu}$.

Representative ringdown trajectories at different positions within the Arnold tongue confirm, that for stronger parametric drives the system becomes increasingly underdamped, see Figs.~\ref{fig:fig3}(d)-(f). For all measurements, we fix the effective detuning at $\mu = -0.2$, yielding a constant angular speed $\tilde \Omega_\mathrm{min} \approx 1.79$. At $\Delta = \SI{15.8}{\hertz}$ and $U_\mathrm{p} = \SI{0.0413}{\volt}$, the system exhibits effective damping $\tilde \gamma \approx 0.544$, see Fig.~\ref{fig:fig3}(d). This trajectory displays clear overdamped signatures: the system relaxes toward the attractor with minimal spiraling. Comparatively, shifting to $\Delta = \SI{36.1}{\hertz}$ and $U_\mathrm{p} = \SI{0.0947}{\volt}$ yields $\tilde \gamma \approx 0.237$, firmly placing the dynamics on the boundary between the over- and underdamped regimes, see Fig.~\ref{fig:fig3}(e). We note that the experimentally accessible drive parameter space intrinsically truncates the deeply underdamped limit: the physical device cannot achieve the necessary $\tilde\gamma \ll 1$ condition required to observe the full Kramers turnover. Notwithstanding, we circumvent this fundamental experimental limitation by complementing our physical measurements with numerical simulations of the slow-flow equations. Namely, we first verify that our numerical integrations perfectly reproduce the experimental trajectories within the accessible parameter window, see Figs.~\ref{fig:fig3}(d)-(e). Subsequently, we leverage the validated simulations to push the system deep into the underdamped regime, reaching $\tilde \gamma \approx 0.0564$, see Fig.~\ref{fig:fig3}(f). This hybrid experimental-numerical methodology enables us to reliably capture the complete turnover spanning the theoretical limits of Eqs.~\eqref{eq_rate__dimless_under} and~\eqref{eq_rate__dimless_over}.

As a third step, we move to study stochastic switching between the phase states of our KPO. To this end, we apply white electrical noise $\xi$ characterized by a standard deviation $\sigma$ (over a bandwidth of 5 MHz) that causes the state of the resonator to fluctuate around its initial phase state~\cite{heugelRoleFluctuationsQuantum2023}. If the fluctuations are large enough, they will occasionally carry the resonator across the threshold between the phase states~\cite{chanPathsFluctuationInduced2008,dykmanFluctuationalPhaseflipTransitions1998,marthalerSwitchingRatesParametric2006,chanActivationBarrierScaling2007,margianiExtractingLifetimeSynthetic2022, hajr2024high, bohm2019understanding, ng2022phaseflip,margianiDeterministicStochasticSampling2022}. The resonator is then captured by the opposite attractor, see inset of Fig.~\ref{fig:fig4}(b). By recording many such events, we can estimate the switching rate $W$~\cite{margianiExtractingLifetimeSynthetic2022}, and rescale it to obtain $\tilde{W}$. In Fig.~\ref{fig:fig4}(a), we show the scaled switching rate $\tilde{W}$ as a function of the effective temperature, see Appendix~\ref{app_calibration} for details.

As a fourth step, in order to observe the KPO's Kramers turnover, we define a trajectory in the phase diagram that possesses constant $\mu$ but varying $\tilde \gamma$, see Fig.~\ref{fig:fig3}(c). Along this trajectory, we measure the scaled switching rate $\tilde{W}$. In order to observe the turnover between the limiting cases in Eqs.~\eqref{eq_rate__dimless_under} and \eqref{eq_rate__dimless_over}, we need to isolate the prefactor from the exponential term. We achieve this in the following way: at each position along the trajectory (each value of $\tilde \gamma$), we estimate $\tilde{W}$ as a function of the effective temperature. As expected, the results approach an exponential dependency, see top row of Fig.~\ref{fig:fig4}(a). We multiply the scaled switching rates by their inverse exponential factors to obtain the scaled prefactor [see bottom row of Fig.~\ref{fig:fig4}(a)], 
\begin{align}
\label{eq_prefactor_KPO_scaled}
	\tilde C &= \tilde W \exp(\tilde \ra /\kb  \tilde{T})\,.
\end{align}
Since analytical expressions for $\tilde \ra$ exist only in limiting cases, we use a numerical approach to find $\tilde \ra$ for arbitrary $\tilde \gamma$, see Appendix \ref{app_switching_KPO} for details.

The prefactors we isolated from the switching rates are expected to cross over from the prefactor $\tilde{C}_\mathrm{u}$ of Eq.~\eqref{eq_rate__dimless_under} to the prefactor $\tilde{C}_\mathrm{o}$ of  Eq.~\eqref{eq_rate__dimless_over}, thereby exposing the KPO's Kramers turnover. In order to show this, we make use of the different temperature dependence of the two prefactors: the prefactor $\tilde{C}_\mathrm{u}$ depends on $1/\tilde T$, while $\tilde{C}_\mathrm{o}$ is independent of $\tilde T$. We express the prefactor at fixed $ \tilde{\gamma}$ as a linear combination of $\tilde{C}_\mathrm{u}$ and $\tilde{C}_\mathrm{o}$,
\begin{align}
	\label{eq_fit_KPO}
	\tilde C(\tilde{T}) = \co \, \frac{(\mu-\mub^{(-)}) |\mub^{(-)}|}{\sqrt{2}\pi \tilde \gamma} + \cu \, \frac{2 \sqrt{1+\mu} I_s(\mu) m }{(2+\mu) k_B \tilde T} \tilde \gamma\,,
\end{align}
and extract the values of $\co$ and $\cu$ from a fit, see bottom row of Fig.~\ref{fig:fig4}(a). In the overdamped limit $\tilde \gamma \approx 1$ \footnote{The boundary of instability is at $\sqrt{1- \mu^2}$. For the chosen value $\mu = 0.2$, this is close to unity.}, we expect $\co \approx 1$ and $\cu \approx 0$, whereas in the underdamped limit $\tilde \gamma \ll 1$ the opposite should hold. 

We show the extracted values of $\co$ and $\cu$ from both experimental and simulated data in Fig.~\ref{fig:fig4}(b). For $\tilde \gamma \sim 1$, we find that $\co$ is close to one, while $\cu$ remains below $0.1$. As $\tilde \gamma$ decreases, the ratio shifts smoothly towards $\cu=\co$ at approximately $\tilde \gamma = 0.15$, after which $\cu$ rises towards one and $\co$ drops below $0.1$. In the overdamped regime where we have both experimental and simulated data, the two methods agree well. Together the two data sets provide strong evidence of a Kramers turnover in a KPO, as predicted by our Eqs.~\eqref{eq_rate__dimless_under} and \eqref{eq_rate__dimless_over}. This is the main result of this work.


In this work, we rely on numerical simulations of the slow-flow equations, Eqs.~\eqref{eq:slow_flow}, to investigate switching dynamics in the deeply underdamped regime. This necessity arises from intrinsic limitations on the maximum drive strength of a KPO before the RWA ceases to be valid. Specifically, the RWA requires the oscillation amplitude to remain sufficiently small such that nonlinear corrections are perturbative, i.e., $\alpha x^2 / 2\omega_0^2 \ll 1$. By explicitly evaluating the steady-state amplitude of the KPO (see Appendix~\ref{app_rescaling}), this constraint can be recast as $\lambda \ll 1$. In contrast, accessing the deeply underdamped regime requires $\tilde{\gamma} \ll \tilde{\Omega}_{\min}$, which translates to $\lambda \gg 2\gamma / \omega_0$. Taken together, these conditions impose the parametric window
\begin{equation}
    2\gamma/\omega_0 \ll \lambda \ll 1.
\end{equation}
For the experimental parameters considered here, $2\gamma / \omega_0 \approx 1.5 \times 10^{-4}$, which significantly restricts the accessible range of $\lambda$ and, consequently, limits the experimental exploration of the deeply underdamped regime.

\section{Conclusion}\label{sec:conclusion}
We have demonstrated a Kramers turnover phenomenon in a KPO. By analyzing the prefactor of the switching rate between the two phase states, we identified a crossover between two distinct dynamical regimes of the rotating-frame dynamics. In the overdamped regime, the prefactor is independent of temperature, whereas in the underdamped regime it acquires a characteristic $1/T$ dependence. By extracting the prefactor from temperature-dependent measurements and numerical simulations, we observe a continuous interpolation between these limits, providing clear evidence for a turnover in the nonequilibrium activation dynamics of a KPO.

On the theoretical side, we derived the analytical form of the prefactor in the underdamped regime, complementing the previously known overdamped result. The prefactor is found to scale linearly with the damping rate and inversely with temperature, in stark contrast to the overdamped case. Revealing this behavior required overcoming two central challenges. First, the physical dissipation rate of the device cannot easily be tuned experimentally. We addressed this by exploiting the driven nature of the KPO: by varying the parametric drive amplitude and frequency, we introduced an effective description in which the conservative part of the dynamics remains fixed while the effective friction is continuously tuned. Second, dissipation in the driven system shifts the attractors in phase space, causing the activation energy itself to depend on the damping and thereby masking the prefactor scaling. We resolved this issue by numerically determining the activation energies, which allowed us to cleanly separate the exponential Arrhenius factor from the prefactor.

Since the turnover does not manifest as a direct maximum of the switching rate, here we identified the turnover through the distinct temperature dependence of the prefactor. This approach provides a practical route to reveal Kramers-type physics in systems where the activation energy varies with the control parameters. While our micro-electromechanical device allowed us to explore the overdamped regime and approach the crossover, the experimentally accessible parameter range does not extend deeply into the underdamped limit. For this reason, we complemented the measurements with numerical simulations of the slow-flow equations, which faithfully reproduce the experimental dynamics and reveal the full turnover. Future experiments using higher-$Q$ devices or platforms with stronger parametric driving could access the deeply underdamped regime directly and observe the entire turnover experimentally.

More broadly, our results establish that the competition between fluctuations and dissipation, first identified by Kramers for equilibrium systems, also governs activation processes in strongly driven nonequilibrium systems. The framework developed here, namely, introducing tunable effective dissipation through parametric driving, provides a controlled route to study activation dynamics in a wide range of platforms, including nanomechanical resonators, superconducting circuits, and quantum parametric devices. More generally, our work highlights that even in complex driven environments, the structure of the activation prefactor encodes fundamental information about the interplay of conservative dynamics, dissipation, and fluctuations.

While significant effort has been devoted to understanding activation mechanisms in the quantum regime \cite{marthalerSwitchingQuantumActivation2006, thompsonQubitDecoherenceSymmetry2022, cardeNonPertQuantum, cortinasUnravelSwitching, OngObserveQuantumHeating, devoretObserveQuantumHeating}, analytical results for the prefactor remain largely restricted to the vicinity of bifurcation points~\cite{marthalerSwitchingQuantumActivation2006, dykmanCriticalExponentsMetastable2007, bonessResonantforceinducedSymmetryBreaking2024}. How a Kramers-like turnover emerges in the quantum domain therefore remains an open question.

\section*{Acknowledgements}
This result constitutes the culmination of approximately six years of continuous experimental and theoretical deliberation. The groups of A.~Eichler and O.~Zilberberg originally initiated the exploration of stochastic turnover physics in the KPO alongside S.~Agrawal, T.~L.~Heugel, and D.~Sabonis. This early collaboration established the preliminary foundation and initial understanding of the transition rates~\cite{agrawal2021transition,agrawal2021transition2}. We thank N.~E.~Bousse and T.~W.~Kenny for providing the MEMS device, and V. Dumont for fruitful discussions. Furthermore, we acknowledge funding from the Deutsche Forschungsgemeinschaft (DFG) through project numbers 449653034, 521530974, and 545605411, as well as via SFB 1432 (project number 425217212). Finally, we acknowledge support from the Swiss National Science Foundation (SNSF) through Sinergia Grant No.~CRSII5\_206008/1. The authors acknowledge support by the local computing resources through the core facility SCCKN.
    
\section*{Appendices}
\setcounter{section}{0}
\renewcommand{\thesection}{\Alph{section}}

\appendix
\section{Notation}
\label{app_notations}
We present a comprehensive summary of the mathematical notation deployed throughout this manuscript. Our work compares the activation dynamics of the static equilibrium double-well potential~\eqref{eq:double_well_pot} with the driven-dissipative KPO~\eqref{eq:KPO_pot}. We rescale both physical models to isolate the underlying conservative dynamics from the environmental coupling. This transformation introduces multiple coordinate mappings and dimensionless parameters, and the resulting proliferation of variables necessitates a consolidated reference. Table~\ref{tab:variables} catalogs the bare physical quantities alongside their rescaled effective counterparts for both models.

\begin{table*}[t!]
\centering
\renewcommand{\arraystretch}{1.5} 
\setlength{\tabcolsep}{0pt} 

\resizebox{\textwidth}{!}{%
\begin{tabular}{| 
    >{\columncolor{mustardBody}\cpad}l<{\cpad} |       
    >{\columncolor{grayABody}\cpad}c<{\cpad} |         
    >{\columncolor{grayBBody}\cpad}c<{\cpad} |         
    >{\columncolor{blueABody}\cpad}c<{\cpad} |          
    >{\columncolor{blueBBody}\cpad}c<{\cpad} |          
}
\cline{2-5}

\multicolumn{1}{l}{} 
 & \multicolumn{2}{|>{\columncolor{grayCHead}\cpad}c<{\cpad}|}{Equilibrium double well} 
 & \multicolumn{2}{|>{\columncolor{blueCHead}\cpad}c<{\cpad}|}{KPO in rotating frame} \\ 
\cline{2-5} 

\multicolumn{1}{l}{} 
 & \multicolumn{1}{|>{\columncolor{grayAHead}\cpad}c<{\cpad}|}{Bare} 
 & \multicolumn{1}{|>{\columncolor{grayBHead}\cpad}c<{\cpad}|}{Rescaled (dimensionless)} 
 & \multicolumn{1}{|>{\columncolor{blueAHead}\cpad}c<{\cpad}|}{Bare} 
 & \multicolumn{1}{|>{\columncolor{blueBHead}\cpad}c<{\cpad}|}{Rescaled (dimensionless)} \\
\hline


Coordinate(s)
& $x$
& $y=\sqrt{a/b}\,x$
& $(u,v)$
& $(Q,P) = \sqrt{-3\alpha/2\lambda \omega_0^2} (u,v)$ \\
\hline

Time
& $t$
& $\tau_\mathrm{k}=\sqrt{b}\,t$
& $t$
& $\tau=\lambda \omega_0^2 t/4\omega_p$ \\
\hline

(Quasi-)Potential
& $V_\mathrm{k}(x)$ \eqref{eq:double_well_pot}
& $V_{\mathrm{k}}(y)$ \eqref{eq_static_pot_scaled}
& $K(u,v)$ \eqref{eq:KPO_pot}
& $g(Q,P)$ \eqref{eq_app_g}\\
\hline

Damping
& $\gamma$
& $\tilde \gamma_{\mathrm{k}}=\gamma/\sqrt{b}$
& $\gamma/2$
& $\tilde \gamma = 2\omega_p \gamma/\lambda \omega_0^2$ \eqref{eq:gamma_eff}\\
\hline

Temperature
& $T$
& $\tilde T_{\mathrm{k}}=a T/b^2$
& $T$
& $\tilde T = -3 \alpha T/2\lambda \omega_0^2 \omega_p^2$ \eqref{eq:T_eff}\\
\hline

Noise force
& $\xi(t)$
& $\xi_y(\tau_\mathrm{k})$
& $\xi_u(t),\xi_v(t)$ \eqref{eq:slow_flow}
& $\xi_Q(\tau),\xi_P(\tau)$ \eqref{eq_app_dynamics_Q}, \eqref{eq_app_dynamics_P} \\
\hline

Noise correlator
& $2\gamma k_B T$
& $2\tilde \gamma_{\mathrm{k}} k_B \tilde T_{\mathrm{k}}$
& $2\gamma k_B T$ \eqref{eq_xi_KPO}
& $2\tilde \gamma k_B \tilde T$ \eqref{eq:xi_KPO_scaled}\\
\hline

Switching rate
& $W_\mathrm{k}$
& $\tilde W_\mathrm{k}$ \eqref{eq_rate_static_scaled}
& $W$ \eqref{eq_equilibrium_rate}
& $\tilde W$ \\
\hline

Prefactor
& $C_\mathrm{k}$ \eqref{eq_equilibrium_rate}
& $\tilde C_{\mathrm{k}}$
& $C$
& $\tilde C$ \eqref{eq_prefactor_KPO_scaled}\\
\hline

Curvature at minimum
& $\omega_\mathrm{min}=\sqrt{2b/m}$
& $\tilde\omega_\mathrm{min}=\sqrt{2}$
& $\Omega_\mathrm{min}=\lambda \omega_0^2 \sqrt{1+\mu}/2\omega_p$ \eqref{eq_omega_min_KPO}
& $\tilde \Omega_\mathrm{min}=2\sqrt{1+\mu}$ \\
\hline

Control parameters
& $a, b$ \eqref{eq:double_well_pot}
& ---
& $\lambda, \omega_\mathrm{p}$ \eqref{eq:KPO_pot}
& $\mu$ \eqref{eq:mu}\\
\hline


\end{tabular}%
}
\caption{\label{tab:variables} \textit{Summary of mathematical notation}. Comparison of bare physical parameters and their rescaled dimensionless counterparts. Left columns detailing the static equilibrium double-well system. Right columns detailing the nonequilibrium KPO within the rotating frame. Equation references provided for explicit mathematical definitions.}

\end{table*}

\section{Kramers Turnover in a Static System}
\label{app_kramers_static}
We provide a thorough discussion of the Kramers turnover in a static system. We first review the principles of thermal activation and the Arrhenius form of the rate. We then detail Kramers' limiting results for the prefactor and their physical implications.

\subsection{Fluctuations and activation energy}
\label{sec: static double well}
We consider a particle of mass $m=1$ moving in a static double-well potential described by Eq.~\eqref{eq:general_eom}, with $V_\mathrm{k} (x) = a x^4/4 - b x^2/2$. To identify the relevant time- and energy-scales governing the dynamics, it is instructive to rescale the equations of motion. We introduce the dimensionless variables $y=\sqrt{a/b}\,x$ and $\tau_\mathrm{k}=\sqrt{b}\,t$. The potential parameters possess dimensions of J/m$^4$ for $a$ and J/m$^2$ for $b$. Hence, the term $\sqrt{b}$ naturally defines a characteristic frequency, whereas $\sqrt{a/b}$ establishes an inverse length scale. Consequently, $y$ and $\tau_\mathrm{k}$ serve as dimensionless quantities that naturally normalize the particle excursion and the characteristic response time of the system.

Substituting these variables into the equation of motion yields the dimensionless form, 
\begin{align}
	\label{eq_static_eom_scaled}
	&\ddot{y} + \partial_y V_{\mathrm{k}}(y) + \tilde{\gamma}_\mathrm{k} \dot{y} = \xi_y(\tau)\,,
\end{align}
perfectly mirroring the rescaling of the RWA equations for the KPO in the main text, cf.~Eqs.~\eqref{eq:gamma_eff} and~\eqref{eq:T_eff} and Appendix.~\ref{app_rescaling}.
The primary advantage of this transformation is that the rescaled potential takes a parameter-free form,
\begin{align}
\label{eq_static_pot_scaled}
V_{\mathrm{k}}(y) = \frac{1}{4}y^4 - \frac{1}{2}y^2\,,
\end{align}
exhibiting two symmetric minima at $y_\pm=\pm1$ with $V_{\mathrm{k},\mathrm{min}}=-1/4$, separated by a barrier at $y=0$ with $V_{\mathrm{k},\mathrm{max}}=0$. The physical complexity of the original problem transfers into two dimensionless control parameters: the effective temperature $\tilde{T}_\mathrm{k} = (a/m b^2)T$ and the effective damping $\tilde{\gamma}_\mathrm{k}=\gamma/\sqrt{b}$, where the former emerges directly from the fluctuation-dissipation relation $\langle \xi_y(\tau)\xi_y(\tau') \rangle = 2\tilde{\gamma}_\mathrm{k} \kb \tilde{T}_\mathrm{k}\,\delta(\tau-\tau')$.

The original physical potential $V(x)$ dictates a local minimum frequency $\omega_{\mathrm{min}} = \sqrt{2b/m}$ and a raw activation barrier $R = b^2/4a$. Mapping these dynamics into our dimensionless framework fundamentally transforms the landscape: the rescaled potential $V_{\mathrm{k}}(y)$ mathematically enforces a constant minimum frequency $\tilde{\omega}_{\mathrm{min}} = \sqrt{2}$ and a fixed dimensionless barrier $\tilde{R} = 1/4$, completely independent of the bare parameters $a$ and $b$. Therefore, the distinction between underdamped and overdamped dynamics is encoded in the effective damping parameter $\tilde\gamma_{\mathrm{k}}$, which evolves with the underlying potential parameters $a$ and $b$. Consequently, transitions between the limiting dynamical regimes can be induced even at fixed bare friction $\gamma$.

Note that varying the bare parameters $a$ and $b$ alters the original physical activation barrier $R$. Hence, observing the Kramers turnover by tuning the potential requires defining a rigid experimental trajectory in the parameter space: the system parameters must be swept simultaneously to continuously vary the effective damping $\tilde\gamma_{\mathrm{k}}$ while keeping the physical barrier $R = b^2/4a$ perfectly constant. Modifying the physical potential landscape along such a constrained trajectory introduces severe experimental challenges: the physical architecture must precisely maintain the exponential Arrhenius argument while continuously deforming the well curvature. 

The rescaling we introduced proposes a valid approach for recent controllable architectures~\cite{Ginot2022}, whereas recent experimental realizations of the Kramers turnover bypassed this approach by maintaining a fixed static potential landscape while directly tuning the bare friction $\gamma$~\cite{rondinDirectMeasurementKramers2017}. In the driven KPO architecture studied here, tuning the physical dissipation rate is experimentally challenging. Instead, here we rely entirely on our rescaling protocol to deploy an analogous constant-barrier trajectory and circumvent this limitation.


Regardless of rescaling, the deterministic limit ($T\to0$) fundamentally suppresses activation: energy dissipation permanently traps the particle within one of the two potential minima. Finite temperatures introduce stochastic forces that successfully propel the particle over the energetic barrier at $x=y=0$, driving escape and subsequent relaxation into the opposing well. This stochastic process of thermal activation exclusively defines the long-time dynamics of the system. We restrict our analysis to the high-barrier limit where such transitions remain rare. There, the switching rate $W_\mathrm{k}$ emerges as the slowest dynamical timescale, separated by several orders of magnitude from the rapid intrawell relaxation rate $\gamma$.

Adopting our rescaled notation, the joint probability density for the position $y$ and momentum $p_y=\dot{y}$ approaches a strict steady state in the limit of thermal equilibrium. Specifically, this density scales proportionally to $\exp\!\left(-[V_{\mathrm{k}}(y) - V_{\mathrm{k},\mathrm{min}} + p_y^2/2]/\kb \tilde{T}_\mathrm{k}\right)$. The resulting switching rate obeys an Arrhenius law, 
\begin{align}
\label{eq_rate_static_scaled}
	\tilde W_\mathrm{k} = \tilde C_\mathrm{k} \exp\!\left[-\frac{V_{\mathrm{k},\mathrm{max}}-V_{\mathrm{k},\mathrm{min}}}{\kb \tilde{T}_\mathrm{k}}\right],
\end{align}
where the prefactor $\tilde C_\mathrm{k}$ depends smoothly on the system parameters. The regime of rare activation is characterized by the condition $(V_{\mathrm{k},\mathrm{max}}-V_{\mathrm{k},\mathrm{min}})/\kb \tilde{T}_\mathrm{k} \gg 1$. As in the main text, we added a tilde to indicate that the rates are in dimensionless time $\tau_\mathrm{k}$.

\subsection{Prefactor and the turnover}
\begin{figure*}[t]
	\centering
	\includegraphics[width=1\linewidth]{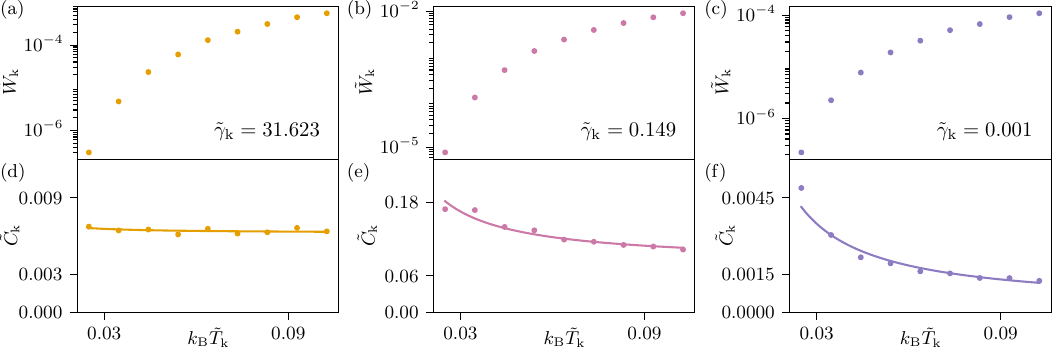}
	\caption{\textit{Switching rate and prefactor fits for the static double-well potential}. 
    (a–c) Scaled switching rate $\tilde{W}_\mathrm{k}$ as a function of $\kb \tilde T$ for different values of $\tilde{\gamma}_\mathrm{k}$, extracted from numerical simulations as described in Appendix~\ref{app_numerical_simulation}.
    (d–f) Rescaled prefactor $\tilde{C}_\mathrm{k}$ extracted from panels (a–c) using Eq.~\eqref{eq_rate_static_scaled}. Solid lines show fits using Eq.~\eqref{eq_fit_static} with $c_\mathrm{u,k}$ and $c_\mathrm{o,k}$ as free parameters. }
	\label{fig_static_turnover_a_b}
\end{figure*}
Kramers originally determined this prefactor by analyzing two asymptotic limiting cases for the effective damping rate $\tilde\gamma_\mathrm{k}$~\cite{kramersBrownianMotionField1940}: the overdamped regime, $\tilde{\gamma}_\mathrm{k} \gg \tilde \omega_\min$, and the underdamped regime, $\tilde{\gamma}_\mathrm{k} \ll  \tilde \omega_\min$. These limits must be  compared with the curvature of the rescaled potential near the minima. Expanding the rescaled potential landscape locally around $y \approx y_\pm$ yields
\begin{align}
	V_{\mathrm{k}}(y) \approx V_{\mathrm{k},\mathrm{min}} + \frac{1}{2}\tilde \omega_\mathrm{min}^2 (y - y_\pm)^2,
\end{align}
with $\tilde \omega_\min=\sqrt{2}$.
Kramers found that the resulting prefactors in the two limiting regimes are given by~\cite{kramersBrownianMotionField1940}
\begin{align}
	\tilde C_\mathrm{o,k} &= \frac{\tilde \omega_\min \tilde \omega_\max}{2\pi \tilde{\gamma}_\mathrm{k}}, \label{eq_static_over}\\
	\tilde C_\mathrm{u,k} &= \frac{1}{2}\frac{\tilde{\gamma}_\mathrm{k}  \tilde S}{k_\mathrm{B}\tilde{T}_\mathrm{k}}\frac{\tilde \omega_\min}{2\pi}, \label{eq_static_under}
\end{align}
where $\tilde \omega_\max$ denotes the effective curvature of the rescaled potential at the barrier top, defined via the expansion
\begin{align}
	V_{\mathrm{k}}(y) \approx -\frac{1}{2} \tilde \omega_\max^2 (y-y_\max)^2,
\end{align}
and $\tilde S$ is the action evaluated at the barrier,
\begin{align}
	\tilde S = 2\int_0^{y_1} \sqrt{-V_{\mathrm{k}}(y)}\,dy,
\end{align}
with the classical turning point $y_1>0$ defined by $U(y_1)=0$. The additional factor of 1/2 appearing in $\tilde C_\mathrm{u,k}$ physically accounts for barrier reflection: an underdamped particle reaching the barrier energy returns to its original well with a probability of 1/2.

The asymptotic expressions of Eqs.~\eqref{eq_static_over} and~\eqref{eq_static_under} mandate a nonmonotonic dependence of the overall switching rate on the damping strength. The rate increases linearly with $\tilde{\gamma}_\mathrm{k}$ in the deep underdamped limit $\tilde{\gamma}_\mathrm{k} \ll \tilde \omega_\min$. Conversely, it    decreases as $1/\tilde{\gamma}_\mathrm{k}$ for $\tilde{\gamma}_\mathrm{k} \gg \tilde \omega_\min$. This dynamic inversion highlights the competing roles of the thermal bath coupling: stochastic fluctuations physically facilitate barrier crossing, whereas viscous dissipation heavily suppresses transport. The prefactor remains temperature independent within the overdamped regime, whereas it scales as $1/\tilde{T}_\mathrm{k}$ in the underdamped limit, see Fig.~\ref{fig:fig2}. We leverage this distinct temperature scaling to isolate and identify the Kramers turnover in systems where the potential can be analytically rescaled.

Mel'nikov analytically resolved the prefactor for arbitrary intermediate damping values~\cite{melnikovKramersProblemFifty1991}, yielding the exact expression
\begin{align}
	&\tilde C_\mathrm{k} = \frac{\tilde \omega_\min}{2\pi}
	\left[\sqrt{1+ \frac{\tilde{\gamma}_\mathrm{k}^2}{4\tilde \omega_\max^2}} - \frac{\tilde{\gamma}_\mathrm{k}}{2\tilde \omega_\max}\right]
	\frac{A(\tilde{\gamma}_\mathrm{k} \tilde S/\tilde{T}_\mathrm{k})^2}{A(2\tilde{\gamma}_\mathrm{k} \tilde S/\tilde{T}_\mathrm{k})},
    \label{eq_melnikov}
\end{align}
with the integral function
\begin{align}
	A(x) = \exp\!\left(\frac{2}{\pi} \int_{0}^{\pi/2}
	\ln\!\left[1-\exp\!\left(-\frac{x}{4\cos^2 z}\right)\right] \mathrm{d}z\right).
\end{align}
This expression smoothly interpolates between the overdamped and underdamped boundaries. It remains valid within the rare-activation limit: the system must satisfy $(V_{\mathrm{k},\mathrm{max}}-V_{\mathrm{k},\mathrm{min}})/\kb \tilde{T}_\mathrm{k} \gg 1$. 

We compare the prefactors extracted from our numerical simulations (see App.~\ref{app_numerical_simulation}) with the full analytical solution of Eq.~\eqref{eq_melnikov} alongside the asymptotic boundaries of Eqs.~\eqref{eq_static_over} and~\eqref{eq_static_under}, cf.~Fig.~\ref{fig:fig2}(a). The overall switching rate exhibits a pronounced maximum at the intermediate coupling $\tilde\gamma_\mathrm{k} \sim \tilde \omega_\min$. This nonmonotonic behavior and appearance of a peak defines the Kramers turnover. We also clearly observe the predicted temperature independence within the overdamped regime. 

\subsection{Kramers turnover by interpolation}
\label{app:Kramers_interpolate}
Kramers turnover can also be identified by analyzing the temperature dependence of the prefactor. To this end, we express the prefactor at fixed $\tilde{\gamma}_\mathrm{k}$ as a linear combination of the two limiting expressions given in Eqs.~\eqref{eq_static_over} and~\eqref{eq_static_under},
\begin{align}
	\label{eq_fit_static}
	\tilde C_\mathrm{k}(\tilde{T}_\mathrm{k}) = c_{\rm o, k}\, \tilde C_\mathrm{o, k} + c_{\rm u, k}\, \tilde C_\mathrm{u ,k}(\tilde{T}_\mathrm{k}).
\end{align}
In the overdamped limit, $\tilde\gamma_\mathrm{k}/ \tilde \omega_\min \gg 1$, we expect $c_{\rm o, k} \approx 1$ and $c_{\rm u, k} \approx 0$, whereas in the underdamped limit $\tilde\gamma_\mathrm{k}/ \tilde \omega_\min \ll 1$ the opposite holds. Both limiting analytical expressions severely overestimate the true physical switching rate within the intermediate crossover regime $\tilde\gamma_\mathrm{k} \sim \tilde \omega_\min$, cf.~Fig.~\ref{fig:fig1}(a). Consequently, the sum $c_{\rm o, k}+c_{\rm u, k}$ drops significantly below unity in this intermediate region.

Using numerical simulations as described in Appendix~\ref{app_numerical_simulation}, we determine the coefficients $c_{\rm o,k}$ and $c_{\rm u,k}$ by fitting $\tilde C(\tilde{T}_\mathrm{k})$ for fixed $\tilde{\gamma}_\mathrm{k}$ as a function of $\tilde{T}_\mathrm{k}$. This procedure is mathematically robust because the overdamped prefactor remains  independent of temperature, whereas the underdamped prefactor scales sharply as $1/\tilde{T}_\mathrm{k}$. In Figs.~\ref{fig_static_turnover_a_b}(a)–(c), we show the results for the rescaled switching rate $\tilde{W}_\mathrm{k}$ as a function of the rescaled temperature. In panels (d)-(f), we show the rescaled prefactor $\tilde C_\mathrm{k}$ together with the fits according to Eq.~\eqref{eq_fit_static}. For $\tilde\gamma_\mathrm{k} \gg \tilde \omega_\min$, see panel (d), the prefactor is almost constant. In contrast, the prefactor exhibits a pronounced temperature dependence, which becomes progressively stronger in panels (e) and (f) as $\tilde \omega_\min$ decreases. The fitted values of $c_{\rm o, k}$ and $c_{\rm u, k}$ are summarized in Fig.~\ref{fig:fig1}(b), where their crossing as a function of $\tilde\gamma_\mathrm{k}$ provides an additional signature of the Kramers turnover.

\section{Numerical simulation of the switching rates}
\label{app_numerical_simulation}
To determine the switching rates, we numerically simulate $N=3000$ stochastic trajectories governed by Eq.~\eqref{eq_static_eom_scaled} for the static system and by Eq.~\eqref{eq_app_dynamics_Q} for the KPO, respectively. All trajectories are initialized at the same stable stationary solution. The analysis begins after a waiting time $\tau_{\mathrm{wait}} \approx 1/\tilde{\gamma}$, ensuring that the system has ``pre-thermalized'' in the vicinity of the initial attractor. Trajectories that have already switched by this time are discarded, leaving \(N_{\mathrm{traj}}<N\) trajectories for the subsequent analysis.

\begin{figure}[t!]
	\centering
	\includegraphics[width=1\linewidth]{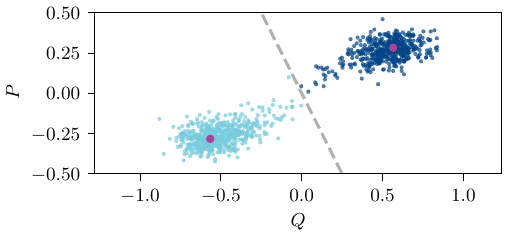}
	\caption{\textit{Stochastic trajectory in rotating phase space}. Stochastic trajectory generated by Eq.~\eqref{eq_app_dynamics_Q} and \eqref{eq_app_dynamics_P} shown at discrete times. In the absence of noise, the system exhibits two stable stationary points (turquoise markers). The gray dashed line, orthogonal to the line connecting these points, defines a separating boundary that assigns each point of the trajectory to one of the two potential wells (light and dark blue). Parameters: $\tilde{\gamma}=0.8$, $\mu=-0.2$, and $\kb \tilde{T}=0.00526$.}
	\label{fig_KPO_numerics_which_well}
\end{figure}

We assign a discrete phase-space identifier $\sigma_i(t_j) \in \{+1,-1\}$ to each remaining trajectory $i$ at every discrete time step $t_j$. This binary value indicates the specific attractor basin currently hosting the trajectory. Specifically, we set $\sigma_i(t_j)=+1$ if the trajectory resides within the initial stable basin, and $\sigma_i(t_j)=-1$ if it has crossed into the opposing basin. Making this dynamic assignment requires introducing a geometric dividing line in phase space: this strict boundary passes precisely through the saddle point and remains  orthogonal to the axis connecting the two attractors, see Fig.~\ref{fig_KPO_numerics_which_well}. This separating line partitions the phase space into two distinct half-planes. If the coordinate of trajectory $i$ at time $t_j$ falls within the initial half-plane, we assign $\sigma_i(t_j)=+1$; otherwise, we assign $\sigma_i(t_j)=-1$.

\begin{figure}[t]
	\centering
	\includegraphics[width=1\linewidth]{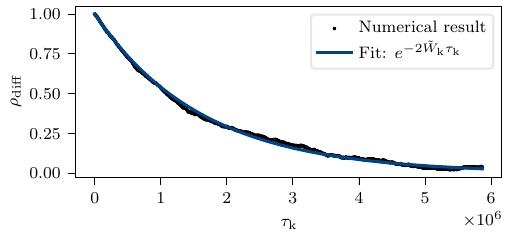}
	\caption{\textit{Extraction of the switching rate from population decay}. Decay of the population imbalance from the numerical simulation and corresponding exponential fit via Eq.~\eqref{eq_app_numerics_fit} treating $\tilde W$ as a free parameter. Example simulation data for a static double-well potential. Parameters: $\kb \tilde{T}_\mathrm{k}=0.025$ and $\tilde{\gamma}_\mathrm{k}=0.0147$.}
	\label{fig_static_numerics_imbalance_decay}
\end{figure}

This binary labeling allows us to estimate the normalized macroscopic populations of the initial and final states,
\begin{align}
\rho_{1}(t_j) &\approx \frac{1}{N_\mathrm{traj}}\sum_{\sigma_i(t_j)=+1}\,,\\\nonumber
\qquad
\rho_{2}(t_j) &\approx \frac{1}{N_\mathrm{traj}}\sum_{\sigma_i(t_j)=-1}\,.
\end{align}
Equivalently, the instantaneous population difference can be written directly in terms of $\sigma_i(t_j)$ as
\begin{align}
\rho_\mathrm{diff}(t_j) &= \rho_1(t_j)-\rho_2(t_j) \approx \frac{1}{N_\mathrm{traj}}\sum_{i=1}^{N_\mathrm{traj}} \sigma_i(t_j).
\end{align}
All trajectories are originally localized near the initial attractor: the initialization protocol enforces $\rho_\mathrm{diff}(0)=1$. As time evolves, stochastic fluctuations drive the trajectories to populate both accessible basins. Therefore, the distinct populations $\rho_1$ and $\rho_2$ asymptotically approach equality, causing $\rho_\mathrm{diff}(t_j)$ to monotonically decrease, see Fig.~\ref{fig_static_numerics_imbalance_decay}.

For rare activation events, the populations $\rho_1$ and $\rho_2$ of the two wells obey a simple balance equation,
\begin{align}
	\dot{\rho}_1 &= W \left(\rho_2 - \rho_1\right),& \dot{\rho}_2 &= -W \left(\rho_2 - \rho_1\right),
\end{align}
where $W$ defines the switching rate between the two wells. The initial condition enforces $\rho_1(0)=1$ and $\rho_2(0)=0$: all simulated trajectories originate identically within a single well. The dynamical population difference $\rho_\mathrm{diff}=\rho_1-\rho_2$ reduces to
\begin{align}
    \dot{\rho}_\mathrm{diff} = -2 W \rho_\mathrm{diff},
\end{align}
and therefore decays exponentially,
\begin{align}
\label{eq_app_numerics_fit}
    \rho_\mathrm{diff}(t) = e^{-2 W t}.
\end{align}

The relation in Eq.~\eqref{eq_app_numerics_fit} provides a direct theoretical link between the numerically obtained population imbalance and the underlying switching rate $W$. We extract $W$ from the simulated datasets by fitting the $\rho_\mathrm{diff}(t)$ traces directly to this exponential form, treating $W$ as the single free fitting parameter, see Fig.~\ref{fig_static_numerics_imbalance_decay}.

We obtain the prefactor $C$ by multiplying the switching rates with the theoretically-expected inverse exponential factor [cf.~Eq.~\eqref{eq:general_rates}] such that
\begin{align}
	C &= W \exp[\tilde \ra/\kb \tilde T].
\end{align}
For the static system, we simply have
\begin{align}
\label{eq_prefactor_static}
	C_\mathrm{k} &= W_\mathrm{k} \exp[(V_{\mathrm{k},\mathrm{max}} - V_{\mathrm{k},\mathrm{min}})/ \kb \tilde T_\mathrm{k}].
\end{align}
For the KPO, we need to numerically find $\ra$ as explained in Appendix~\ref{app_switching_KPO}.

We rescale the extracted bare prefactor $C$ to its dimensionless counterpart $\tilde{C}$. This transformation follows our established mapping protocols for both the static equilibrium system and the driven KPO, cf. Appendix~\ref{app_kramers_static} and~\ref{app_rescaling}. The resulting dimensionless prefactor $\tilde{C}$ manifests explicitly as a function of the effective temperature $\tilde{T}$. We evaluate this thermal response by repeatedly extracting the switching dynamics across a broad range of noise strengths: the underlying physical control parameters are held constant at fixed $a$ and $b$ for the static case, and fixed $\lambda$ and $\omegap$ for the KPO, and we increase the amplitude of $\xi$. We subsequently fit the empirical temperature-dependent profile $\tilde{C}(\tilde{T})$ using a linear superposition of the analytical overdamped and underdamped asymptotic boundaries, cf.~Eqs.~\eqref{eq_fit_static} and~\eqref{eq_fit_KPO} for static a KPO cases, respectively. This analytical projection cleanly isolates the phenomenological crossover coefficients $c_{\mathrm{o}}$ and $c_{\mathrm{u}}$. Tracking the evolution of these isolated coefficients directly against the varying effective damping $\tilde{\gamma}$ successfully exposes the fundamental Kramers turnover, see Figs.~\ref{fig:fig2} and~\ref{fig:fig4}.

\section{Scaling of the RWA dynamics}
\label{app_rescaling}
We rescale the rotating-frame quadratures $(u,v)$ and the physical time $t$ to derive an effective friction coefficient governed by the parametric modulation amplitude. Specifically, we introduce the dimensionless transformations
\begin{align}
	\tau &= \lambda \omega_0^2 t/4\omegap\,, &	(u,v) &= \sqrt{-\frac{2\lambda \omega_0^2}{3\alpha}}(-P,Q)\,.
\end{align}
Applying these mappings to the slow-flow dynamics~\eqref{eq:slow_flow} yields the dimensionless equations of motion,
\begin{align}
	&\frac{\mathrm{d} Q}{\mathrm{d} \tau} = \frac{\partial g}{\partial P} - \tilde \gamma Q + \xi_Q(\tau) \,, \label{eq_app_dynamics_Q}\\
	&\frac{\mathrm{d} P}{\mathrm{d} \tau} = -\frac{\partial g}{\partial Q} - \tilde \gamma P + \xi_P(\tau)\,, \label{eq_app_dynamics_P}
\end{align}
where the effective RWA quasipotential $g(Q,P)$ and the stochastic noise correlators are defined as: 
\begin{align}
    \label{eq_app_g}
	&g(Q,P) = \frac{1}{4}(Q^2+P^2)^2 + \frac{1}{2}(1-\mu) P^2 - \frac{1}{2}(1+\mu) Q^2\,,~ \\
	&\langle \xi_i(\tau) \xi_j(\tau') \rangle =  2 \delta(\tau-\tau') \delta_{ij} \tilde \gamma \kb \tilde T/m \,.
    \label{eq:xi_KPO_scaled}
    \end{align}
For convenience, we repeat that $\tilde \gamma = 2\omegap \gamma/\lambda \omega_0^2$ and $\tilde T = -3 \alpha T/2 \lambda \omega_0^2 \omegap^2$, as defined in the main text. Crucially, the conservative quasipotential $g(Q,P)$ governing the deterministic evolution depends on a single dimensionless parameter: the rescaled detuning $\mu = 2 (\omega_0^2 - \omegap^2) /\lambda \omega_0^2$. Hence, simultaneously tuning the physical drive frequency $\omegap$ and amplitude $\lambda$ allows this conservative landscape to remain invariant.

\begin{figure}[t]
	\centering
	\includegraphics[width=1\linewidth]{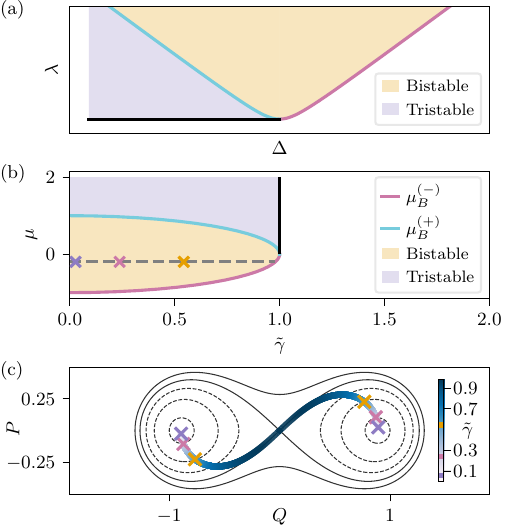}
	\caption{\textit{Instability lobe and attractor structure of the  KPO}. 
    (a) Instability lobe of the KPO as a function of detuning $\Delta$ and modulation amplitude $\lambda$. 
    (b) Same instability lobe expressed in terms of the scaled parameters $\tilde{\gamma}$ and $\mu$. The gray dashed line indicates the cut at $\mu=-0.2$, along which all measurements and simulations were performed. 
    (c) Phase-space position of the two attractors versus $\tilde{\gamma}$ for $\mu=-0.2$. Solid and dashed lines denote contour lines of $g(Q,P)$ [cf. Eq.~\eqref{eq_app_g}] for values above and below the saddle point at $g=0$, respectively. Crosses mark the parameter values at which the ringdown experiments and simulations discussed in the main text were taken (cf.~Fig.~\ref{fig:fig3}), i.e., at $\tilde{\gamma}=0.543, 0.237, 0.056$.}
	\label{fig_app_lobe}
\end{figure}

The stable stationary solutions of the deterministic equations [cf.~Eqs.~\eqref{eq_app_dynamics_Q} and~\eqref{eq_app_dynamics_P}] define the states of forced parametric vibration. Depending on the parameters $\mu$ and $\tilde \gamma$ there may be one, two, or three such attractors, see Fig.~\ref{fig_app_lobe}. In the regime where there is only one attractor, we have only the trivial solution $Q=P=0$, which we dub the non-oscillating regime. The other two regimes, where attractors at finite values of $Q$ and $P$ appear, are referred to as the bistable and tristable regimes, respectively.

The boundaries separating these distinct dynamical phases correspond to bifurcations of the underlying equations of motion. Specifically, the two critical bifurcation lines isolating the bistable regime from both the non-oscillating and tristable phases are analytically defined in the rescaled parameter space by:
\begin{align}
    \mub^{(\mp)} = \mp \sqrt{1-\tilde\gamma^2}\,.
\end{align}
We confine our analysis to the bistable regime bounded by $\mub^{(-)}< \mu < \mub^{(+)}$. This region hosts two stable attractors separated by a single central saddle point. Unlike the static equilibrium system (cf.~Appendix~\ref{app_kramers_static}), the coordinates of these dynamic attractors shift as a function of the effective damping $\tilde \gamma$, see Fig.~\ref{fig_app_lobe}(c). The radial distance of these attractors from the phase-space origin physically defines the amplitude of the forced vibrations. Mapping back to the physical variables yields the dimensionful steady-state amplitude,
%
\begin{align}
\label{eq_amplitude_KPO}
    A = \sqrt{-\frac{2 \lambda \omega_0^2}{3 \alpha}} \sqrt{\sqrt{1-\tilde \gamma^2} + \mu}.
\end{align}
The location, $Q=P=0$, of the unstable solution is independent of $\tilde \gamma$.


%

\section{Switching rates of the KPO}
\label{app_switching_KPO}
Finding the switching rates for the KPO introduces significant theoretical complexity compared with the static double-well potential. This difficulty arises fundamentally because the KPO operates as a driven out-of-equilibrium system and detailed balance is violated.

In the regime where the noise intensity is small, such that switching between the two attractors is rare, one can find the switching rate $W$ to logarithmic accuracy by looking at the least improbable realization of the stochastic force that results in a transition. This path can be found by solving a variational problem~\cite{freidlinRandomPerturbationsDynamical}, by looking at an auxiliary conservative system described by the Lagrangian
\begin{align}
	&\mathcal{L}(\dot{\mathbf{q}}, \mathbf{q}) = \frac{1}{2} \left[\dot{\mathbf{q}} - \mathbf{K}(\mathbf{q}) \right]^2\,,
\end{align}
where the generalized coordinates and force vectors are defined as
\begin{align}
&\mathbf{q} = \left(\begin{matrix}
		Q \\ P
	\end{matrix}\right),\qquad \rm{and } \qquad \mathbf{K}(\mathbf{q}) = \left( \begin{matrix}
		\partial_P g(Q,P) - \tilde \gamma Q\\
		-\partial_Q g(Q,P) - \tilde \gamma P		
	\end{matrix}\right).\nonumber
    \end{align}
The dimensionless activation energy $\tilde \ra$ emerges from the minimization of the classical action,
\begin{align}
	\tilde \ra = \frac{1}{2\tilde\gamma} \mathrm{min} \int_{-\infty}^{\infty} \mathrm{d} \tau  \mathcal{L}(\dot{\mathbf{q}}, \mathbf{q})\,.
\end{align}
This minimization is constrained to trajectories originating at a stable stationary point at $\tau \to - \infty$ and terminating at the central saddle point $\mathbf{q} = (0,0)$ as $\tau \to \infty$. 
\begin{figure}[t]
	\centering
	\includegraphics[width=1\linewidth]{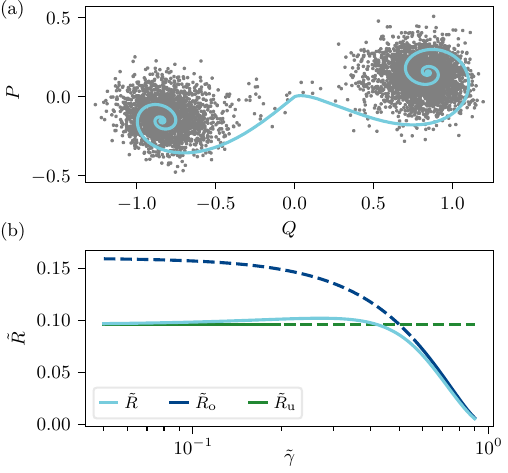}
	\caption{\textit{Optimal escape path and activation exponent}. (a) Optimal escape path (solid line) originating from the $Q>0$ attractor for $\mu=-0.2$ and $\tilde{\gamma}=0.3$. Sampled stochastic trajectory points (dots) generated via Eqs.~\eqref{eq_app_dynamics_Q} and~\eqref{eq_app_dynamics_P} undergoing multiple switching events. 
    (b) Numerically calculated activation exponent $\tilde{R}_\mathrm{a}$ compared with the overdamped and underdamped approximations [cf.~Eqs.~\eqref{eq_app_ra_over} and~\eqref{eq_app_ra_under}].}
	\label{fig_freidlin_wentzel}
\end{figure}
In Fig.~\ref{fig_freidlin_wentzel}(a), we show an example of an optimal escape path obtained using \textit{CriticalTransitions.jl}~\cite{heugel2024,CriticalTransitionsjl}. In Fig.~\ref{fig_freidlin_wentzel}(b), we show the dimensionless activation energy $\tilde \ra$ as a function of $\tilde\gamma$ for $\mu=-0.2$, compared with the analytical asymptotic expressions~\cite{dykmanFluctuationalPhaseflipTransitions1998}. The dimensionless activation energy $\tilde \ra$ dictates the dominant exponential scaling of the rescaled switching rate $\tilde W$. We isolate this exponential argument via finding the optimal escape path. In the following, we proceed to evaluate the full macroscopic transition rate: we analytically derive the explicit structural prefactors governing the asymptotic overdamped and underdamped limiting regimes.

Two distinct deterministic contributions dictate the dynamics of the rescaled quadratures, cf.~Eqs.~\eqref{eq_app_dynamics_Q} and~\eqref{eq_app_dynamics_P}: the conservative force originating from the quasipotential $g(Q,P)$ and the dissipative force scaling with $\tilde \gamma$. The conservative force acting in isolation generates perfectly periodic orbits of constant energy $g(Q,P)=g$ characterized by a frequency $\tilde\Omega(g)$. Conversely, the dissipative force structurally breaks this periodicity: it collapses specific stationary solutions into stable attractors. A separation of timescales emerges within two asymptotic limits, explicitly defining a slow and a fast evolving coordinate. Exploiting this separation reduces the complex phase-space dynamics to a single effective degree of freedom~\cite{dykmanFluctuationalPhaseflipTransitions1998, dykmanScalingActivatedEscape2005}. This dimensional reduction yields explicit analytic solutions for the macroscopic switching rates, providing closed-form expressions for both the prefactors and the activation energies.

\subsection{The overdamped regime}
\label{app_rates_over}
The standard overdamped regime of a generic harmonic oscillator is mathematically defined by a strict condition: the dissipation rate $\gamma$ must exceed twice the natural eigenfrequency $2\omega_0$. Thus, the time derivative of the momentum becomes negligible, instantly collapsing the dynamics to a single spatial dimension. 

For the bistable KPO, it turns out that $\tilde \gamma$ cannot become larger than $\tilde \Omega_\min$,  prohibiting a conventional overdamped limit. Notwithstanding, a slow-fast coordinate separation emerges precisely as the system approaches the bifurcation into the non-oscillating regime~\cite{dykmanFluctuationalPhaseflipTransitions1998, dykmanScalingActivatedEscape2005}. To see this, we first define a small detuning parameter $\epsilon = \mu-\mub^{(-)}$ to define dynamics within the bistable regime that can get arbitrarily close to the critical bifurcation boundary ($0< \epsilon \ll 1$). Note that for a fixed $\mu$, $\epsilon$ is a only a function of $\tilde{\gamma}$, leading to an evolution of the locations of the phase states as depicted in Fig.~\ref{fig_app_lobe}(c).
Correspondingly, we rotate the coordinates $(Q,P)$ by an angle $\delta$ towards the new locations of the minima by 
\begin{align}
	\sin 2\delta &= \tilde \gamma\,, & \cos 2\delta &= \mub^{(-)}\,,
\end{align}
such that
\begin{align}
    \left(\begin{matrix}
        \tilde Q\\ \tilde P
    \end{matrix}\right) &= \left(\begin{matrix}
        \cos \delta & \sin \delta\\
        -\sin \delta & \cos \delta
    \end{matrix}\right) \left(\begin{matrix}
        Q\\ P
    \end{matrix}\right).
\end{align}
In this rotated system, we find that $\tilde{P}$ evolves by a factor $\epsilon$ faster in time than $\tilde{Q}$~\cite{dykmanFluctuationalPhaseflipTransitions1998, dykmanScalingActivatedEscape2005}. Hence, we can average over the fast degree of freedom, $\tilde{P}$, and reduce the dynamics to a single coordinate
\begin{align}
	\frac{\mathrm{d} \tilde Q}{\mathrm{d} \tau} &= - \frac{\partial V_\eff(\tilde Q)}{\partial \tilde Q} + \xi_{\tilde Q}(\tau)\,,
\end{align}
where the noise term is characterized by $\langle \xi_{\tilde Q}(\tau) \xi_{\tilde Q}(\tau') \rangle =  2 \tilde\gamma \kb \tilde T \delta (\tau-\tau')$, and the potential
\begin{align}
	V_\eff(\tilde Q) &= \frac{1}{4}\frac{|\mub^{(-)}|}{\tilde \gamma^3} {\tilde Q}^4 - \frac{1}{2} \epsilon \frac{|\mub^{(-)}|}{\tilde \gamma} {\tilde Q}^2\,,
\end{align}
defines a one-dimensional static double well. For such a one-dimensional stochastic problem, the associated Fokker-Planck equation is a Smoluchowski equation~\cite{kramersBrownianMotionField1940}. Hence, in this limit, the switching rates between the states of forced vibration perfectly map to the classical escape rates between these two effective static minima, cf.~Appendix~\ref{sec: static double well}. We can calculate their asymptotic limits by directly utilizing the original Kramers expressions~\cite{kramersBrownianMotionField1940}. Cast in dimensionless time, the overdamped switching rate and its corresponding activation energy therefore read~\cite{dykmanFluctuationalPhaseflipTransitions1998, dykmanScalingActivatedEscape2005}
\begin{align}
	\tilde W_\over &= \frac{\epsilon |\mub^{(-)}|}{\sqrt{2}\pi \tilde \gamma} \exp \left[-\frac{\epsilon^2 |\mub^{(-)}|}{4 \kb \tilde T}\right]\,,\\
	\tilde \ra_\over &= \epsilon^2  |\mub^{(-)}|/4\,.
	\label{eq_app_ra_over}
\end{align}

\subsection{The underdamped regime}
\label{app_rates_under}
For sufficiently large drives where the system is far from the critical bifurcation boundary $\mub^{(-)}$, the conservative rotating-frame dynamics dominates the effective friction $\Omega_\min \gg \tilde \gamma$ in the motion around the phase space attractors, which we call the underdamped regime. 
 
From Eqs.~\eqref{eq_app_dynamics_Q} and~\eqref{eq_app_dynamics_P}, we can directly write down the corresponding Fokker-Planck equation for the probability density $\rho$ \cite{riskenFokkerPlanckEquationMethods2012}:
\begin{align}
    \frac{\mathrm{d} \rho}{\mathrm{d} \tau} = &-\partial_Q \left[ \left( \frac{\partial g}{\partial P}-\tilde \gamma Q\right)\rho\right]-\partial_P \left[ \left( -\frac{\partial g}{\partial Q}-\tilde \gamma P\right)\rho\right] \nonumber \\
    &+ \tilde \gamma \kb \tilde T (\partial_Q^2 + \partial_P^2) \rho\,.
    \label{eq:FP_underdamped_KPO}
\end{align}
In the deep underdamped limit, particles move approximately in closed orbits in phase space with a given energy $g$. As such, the stochastic dynamics are better described as transitions between different value of energy, and not explicitly on specific locations in $Q$ and $P$, i.e., we can write $\rho(Q,P)\rightarrow \rho(g)$. Hence, using the chain rule, we can write the Fokker-Planck equation~\eqref{eq:FP_underdamped_KPO} in the form
\begin{align}
    \frac{\mathrm{d} \rho}{\mathrm{d} \tau} = 2\tilde \gamma \rho &+ \tilde \gamma \left(Q\frac{\partial g}{\partial Q} +  P \frac{\partial g}{\partial P}\right) \frac{\partial \rho}{\partial g}\nonumber \\ 
    &+ \tilde \gamma \kb \tilde T \left( \frac{\partial^2 g}{\partial Q^2} + \frac{\partial^2 g}{\partial P^2}\right) \frac{\partial \rho}{\partial g}\nonumber \\
    &+ \tilde \gamma \kb \tilde T \left[\left(\frac{\partial g}{\partial Q}\right)^2 + \left(\frac{\partial g}{\partial P}\right)^2 \right] \frac{\partial^2 \rho}{\partial g^2}\,.
\end{align}
We assume that the deterministic dynamics along a closed orbit with a given $g$ is much fast than the weak stochastic kicks between orbits and much faster than dissipation (underdamped). Averaging over the fast oscillations in the rotating frame, we find a diffusion equation for the probability distribution \cite{dmitrievActivatedTunnelingTransitions1986}
\begin{align}
    \label{eq_FP_action}
	\frac{\mathrm{d} \rho(I)}{\mathrm{d} \tau} &= 2 \tilde \gamma \partial_I \left[I \rho(I) + \kb \tilde T N(I) \frac{\partial \rho(I)}{\partial g} \right]\,,
\end{align}
that depends on the action $I$ and the diffusion coefficient $N$,
\begin{align}
	&I(g) = \frac{1}{2\pi} \iint_{A(g)} \mathrm{d} Q \mathrm{d}P \,, \\
	&N(g) = \frac{1}{4\pi} \iint_{A(g)} \nabla^2 g(Q,P) \mathrm{d} Q \mathrm{d}P\,.
\end{align}
The double integrals run over the interior $A(g)$ of the well with $Q>0$. Note that $I$, $N$, and $\tilde \Omega_\min$ depend on $\mu$. 

With Eq.~\eqref{eq_FP_action} at hand, we can follow Kramers original work for the static underdamped double well~\cite{kramersBrownianMotionField1940} to find the switching rate. Specifically, we can write down the continuity equation
\begin{align}
    \frac{\mathrm{d} \rho}{\mathrm{d} \tau} &= - \partial_I w_I\,,
\end{align}
where we defined the current
\begin{align}
	&w_I = -2 \tilde \gamma \kb \tilde T N(I) e^{\frac{-\tilde R(g)}{\kb \tilde T}} \partial_g \left[e^{\frac{\tilde R(g)}{\kb \tilde T}} \rho (I)\right]\,, 
\end{align}
with the dimensionless $g$ dependent barrier 
\begin{align}
	&\tilde R(g) = \int_{g_{\mathrm{min}}}^g \frac{I(g')}{N(g')} \mathrm{d} g'\,,
\end{align}
and $g_{\mathrm{min}}$ is the value of $g$ at the bottom of the well.

By integrating from $g$ to the saddle point $g_\mathrm{s} = 0$ and introducing $I_s(\mu) = I(g_s,\mu)$, we find
\begin{align}
	\frac{\rho(I)}{w_I} &\approx \frac{1}{2\tilde \gamma I_\mathrm{s} (\mu)} e^{\frac{\tilde R(g_\mathrm{s})}{\kb \tilde T}} e^{\frac{-\tilde R(g)}{\kb \tilde T}}\,.
\end{align}
The rate at which particles reach the barrier is then given by $w/n_0$, where $n_0 = \int \rho(I) \mathrm{d}I$. In the energy diffusion regime, discussed here, the probability that the particle relaxes towards either well is $1/2$. Hence, the switching rate has an additional factor of $1/2$. With that 
\begin{align}
	\tilde W_\under &= \frac{\tilde \Omega_\min I_\mathrm{s}(\mu)}{(2+\mu) \kb \tilde T} \tilde\gamma \exp \left[-\frac{\tilde \ra_\under}{\kb \tilde T}\right]\,, \nonumber \\
	\tilde \ra_\under &= \tilde R(g_\mathrm{s})\,.
		\label{eq_app_ra_under}
\end{align}
The expression for the prefactor derived here is one of the main results of this work.

\section{Experimental calibration and uncertainty analysis}
\label{app_calibration}
In this section, we describe how the parameters of the experimental setup are determined and how their uncertainties are propagated through the data analysis.

\subsection{Parameter extraction}
The resonance frequency and decay rate of the resonator are obtained by weakly driving the device with an external tone and fitting the measured response with a Lorentzian line shape. From this procedure we obtain a resonance frequency 
$\omega_0/(2\pi) = \SI{1.13}{\mega\hertz}$ and a decay rate $\gamma = \SI{537}{\per\second}$.

The conversion between the applied modulation voltage $U_\mathrm{p}$ and the parametric drive amplitude $\lambda$ is given by
\begin{align}
    \lambda = \frac{2 U_\mathrm{p}}{Q U_\mathrm{th}},
\end{align}
where $U_\mathrm{th}=\SI{0.02245}{\volt}$ denotes the parametric threshold voltage and 
$\omega_0/\gamma = 13278$ is the quality factor of the resonator.

The nonlinear coefficient $\alpha$ (expressed in voltage units) is extracted by fitting the measured steady-state amplitude of the KPO as a function of detuning to the theoretical prediction, Eq.~\eqref{eq_amplitude_KPO}. From this fit we obtain
\begin{align}
    \alpha = \SI{-4.43e16}{\volt^{-2}\second^{-2}} .
\end{align}

\subsection{Noise calibration}
To relate the applied electrical noise to an effective temperature, we compare the experimentally measured switching rates $W$ with numerical simulations of the stochastic slow-flow equations. This procedure yields the equipartition relation
\begin{align}
    k_\mathrm{B}T = a \sigma^2 ,
\end{align}
where $\sigma$ is the standard deviation of the applied voltage noise and 
$a = \SI{3e6}{\coulomb\per\volt}$ is the calibration factor.

\subsection{Uncertainty propagation}
Several experimentally determined parameters carry uncertainties that must be taken into account in the analysis. Table~\ref{tab:parameters} summarizes the values used in the main text together with their estimated uncertainties. The quoted uncertainty of $f_0$ corresponds to the uncertainty of the detuning between the modulation frequency and the instantaneous resonance frequency, after correcting for slow frequency drifts of the resonator.
\begin{table}[h!]
\centering
\begin{tabular}{|c|c|c|}
\hline
Parameter & Value & Uncertainty \\
\hline
$f_0$ & \SI{1.13}{\mega\hertz} & \SI{1}{\hertz} \\
$\gamma$ & \SI{537}{\per\second} & \SI{0.5}{\per\second} \\
$U_\mathrm{th}$ & \SI{0.02245}{\volt} & \SI{0.0003}{\volt} \\
$\alpha$ & \SI{-4.43e16}{\volt^{-2}\second^{-2}} & \SI{1.34e14}{\volt^{-2}\second^{-2}} \\
$a$ & \SI{3e6}{\second^{-2}} & \SI{0.05e6}{\second^{-2}} \\
\hline
\end{tabular}
\caption{Experimental parameters and their uncertainties.}
\label{tab:parameters}
\end{table}

To propagate these uncertainties through the data analysis, we vary each parameter within its uncertainty range, repeat the full analysis pipeline, and evaluate the resulting variation of the extracted quantities. This procedure yields the uncertainties of the coefficients $c_\mathrm{o}$ and $c_\mathrm{u}$ shown in Fig.~\ref{fig:fig4}.


\bibliography{zotero_daniel}
\end{document}